\documentclass[12pt,preprint]{aastex}

\slugcomment{Not to appear in Nonlearned J., 45.}

\shorttitle{UV to IR SEDs of UV selected galaxies in the ELAIS fields}
\shortauthors{Iglesias-P\'{a}ramo et al.}

\begin{document}


\title{UV to IR SEDs of UV selected galaxies in the ELAIS fields: evolution of dust attenuation and star formation activity from $z=0.7$ to $z=0.2$}


\author{J. Iglesias-P\'{a}ramo\altaffilmark{1}, 
V. Buat\altaffilmark{2}, 
J. Hern\'{a}ndez-Fern\'{a}ndez\altaffilmark{1}, 
C. K. Xu\altaffilmark{3,4}, 
D. Burgarella\altaffilmark{2}, 
T. T. Takeuchi\altaffilmark{5}, 
A. Boselli\altaffilmark{2}, 
D. Shupe\altaffilmark{6}, 
M. Rowan-Robinson\altaffilmark{7}, 
T. Babbedge\altaffilmark{7}, 
T. Conrow\altaffilmark{3}, 
F. Fang\altaffilmark{6}, 
D. Farrah\altaffilmark{8},
E. Gonz\'{a}lez-Solares\altaffilmark{9}, 
C. Lonsdale\altaffilmark{4}, 
G. Smith\altaffilmark{10}, 
J. Surace\altaffilmark{6}, 
T. A. Barlow\altaffilmark{3}, 
K. Forster\altaffilmark{3}, 
P. G. Friedman\altaffilmark{3}, 
D. C. Martin\altaffilmark{3}, 
P. Morrissey\altaffilmark{3}, 
S. G. Neff\altaffilmark{11}, 
D. Schiminovich\altaffilmark{12}, 
M. Seibert\altaffilmark{3}, 
T. Small\altaffilmark{3}, 
T. K. Wyder\altaffilmark{3}, 
L. Bianchi\altaffilmark{13}, 
J. Donas\altaffilmark{2}, 
T. M. Heckman\altaffilmark{14}, 
Y.-W. Lee\altaffilmark{15}, 
B. F. Madore\altaffilmark{16}, 
B. Milliard\altaffilmark{2}, 
R. M. Rich\altaffilmark{17}, 
A. S. Szalay\altaffilmark{14}, 
B. Y. Welsh\altaffilmark{18} and 
S. K. Yi\altaffilmark{15}}

\altaffiltext{1}{Instituto de Astrof\'{\i}sica de Andaluc\'{\i}a, Camino Bajo de Hu\'{e}tor 50, 18008 Granada, SPAIN}
\altaffiltext{2}{Observatoire Astronomique Marseille Provence, Laboratoire d'Astrophysique de Marseille, 13012, Marseille, FRANCE}
\altaffiltext{3}{California Institute of Technology, MC 405-47, 1200 East California Boulevard, Pasadena, CA 91125}
\altaffiltext{4}{Infrared Processing and Analysis Center, California Institute of Technology, 100-22, Pasadena, CA91125}
\altaffiltext{5}{Institute for Advanced Research, Nagoya University, Furo-cho, Chikusa-ku, Nagoya 464-8601, JAPAN}
\altaffiltext{6}{Spitzer Science Center, California Institute of Technology, Mail Stop 220-6, Pasadena, CA 91125}
\altaffiltext{7}{Astrophysics Group, Blackett Laboratory, Imperial College of Science, Technology and Medicine, Prince Consort Road, London SW7 2BZ, UK}
\altaffiltext{8}{Department of Astronomy, Cornell University, Ithaca, NY 14853}
\altaffiltext{9}{Institute of Astronomy, Madingley Road, Cambridge CB3 0HA, UK}
\altaffiltext{10}{Center for Astrophysics and Space Sciences, University of California, San Diego, La Jolla, CA 92093-0424}
\altaffiltext{11}{Laboratory for Astronomy and Solar Physics, NASA Goddard Space Flight Center, Greenbelt, MD 20771}
\altaffiltext{12}{Department of Astronomy, Columbia University, New York, NY 10027}
\altaffiltext{13}{Center for Astrophysical Sciences, The Johns Hopkins University, 3400 N. Charles St., Baltimore, MD 21218}
\altaffiltext{14}{Department of Physics and Astronomy, The Johns Hopkins University, Homewood Campus, Baltimore, MD 21218}
\altaffiltext{15}{Center for Space Astrophysics, Yonsei University, Seoul 120-749, Korea}
\altaffiltext{16}{Observatories of the Carnegie Institution of Washington, 813 Santa Barbara St., Pasadena, CA 91101}
\altaffiltext{17}{Department of Physics and Astronomy, University of California, Los Angeles, CA 90095}
\altaffiltext{18}{Space Sciences Laboratory, University of California at Berkeley, 601 Campbell Hall, Berkeley, CA 94720}




\begin{abstract}
We study the ultraviolet to far-infrared (hereafter UV-to-IR) SEDs of a sample of intermediate redshift ($0.2 \leq z \leq 0.7$) UV-selected galaxies from the ELAIS-N1 and ELAIS-N2 fields by fitting a multi-wavelength dataset to a library of GRASIL templates. 
Star formation related properties of the galaxies are derived from the library of models by using the Bayesian statistics. 
We find a decreasing presence of galaxies with low attenuation and low total luminosity as redshift decreases, which does not hold for high total luminosity galaxies.
In addition the dust attenuation of low mass galaxies increases as redshift decreases, and this trend seems to disappear for galaxies with $M* \geq 10^{11}$~$M_{\odot}$. 
This result is consistent with a mass dependent evolution of the dust to gas ratio, which could be driven by a mass dependent efficiency of star formation in star forming galaxies.
The specific star formation rates (SSFR) decrease with increasing stellar mass at all redshifts, and for a given stellar mass the SSFR decreases with decreasing redshift. The differences in the slope of the $M*$--SSFR relation found between this work and others at similar redshift could be explained by the adopted selection criteria of the samples which, for a UV selected sample, favours blue, star forming galaxies.

\end{abstract}





\section{Introduction}

Multi-band observations of galaxies are becoming increasingly important for a coherent understanding of the formation and evolution of galaxies. 
The increasing availability of wide field surveys at different wavelengths (e.g. GALEX, Spitzer, 2MASS, SDSS, 2dF) provides a wealth of data to better constrain the evolution of galaxies by analyzing it from several physical standpoints. 
In the last decade, wide area surveys have evidenced a strong evolution of some galaxy properties with redshift, particularly those related to star formation and dust attenuation. 
In particular, the local star formation rate density was found to be far lower than it was at $z \approx 1$ (Lilly et al. 1996; Madau et al. 1996; Flores et al. 1999; Hopkins 2004), although there is no clear agreement on the epoch where it was maximum (Heavens et al. 2004).
A strong evolution of luminosities with redshift is also suggested by the study of the rest frame ultraviolet (UV; Arnouts et al. 2005) and rest frame infrared (IR; Xu 2000; Chary \& Elbaz 2001; Le Floc'h et al. 2005) luminosity functions at different redshifts.

The overall dust attenuation in star forming galaxies also shows signatures of evolution with redshift. 
Takeuchi et al. (2005a) have reported a strong decrease in the ratio of the IR to UV star formation rate (SFR) densities ($\rho_{IR} / \rho_{UV}$) between $z = 1$ and $z = 0$, which can be interpreted as a decrease over this redshift range in the dust attenuation of galaxies averaged over a volume element.
The situation is far from clear when individual galaxies are studied.
Buat et al. (2007) and Xu et al. (2007) report a slightly lower dust attenuation for LIRGs at $z \simeq 0.6-0.7$ compared to LIRGs at $z = 0$ which seems not to be true for ULIRGs (Xu et al. 2007). 
Burgarella et al. (2007) reported a lower dust attenuation (as measured from the IR/UV ratio) for LBGs at $z \simeq 1$, compared to the sample of UV selected local galaxies of Buat et al. (2006), and 
these LBGs are more attenuated than the sample of Reddy et al. (2006) at $z \simeq 2$. 

Another quantity which has been extensively studied is the specific SFR (SSFR), which is defined as the ratio of the SFR to the stellar mass of the galaxy. This SSFR is now used to study the downsizing phenomenon -- i.e. when $z$ decreases intense star formation activity becomes limited to less and less massive galaxies --  from $z \simeq 5$ to present time (e.g. Cowie et al. 1996; Heavens et al. 2004; Brinchmann \& Ellis 2000; Boselli et al. 2001; Bauer et al. 2005; Feulner et al. 2005). 
High mass galaxies with values of SSFR as high as those of low mass galaxies have been reported in the local Universe, most of them being LIRGs and/or ULIRGs (Hammer et al. 2005; Iglesias-P\'{a}ramo et al. 2006; Buat et al. 2006), but they are far less numerous than the massive galaxies with negligible star formation activity. 
Another interesting result reported by Bell et al. (2005) is that the SSFR of high mass galaxies was much higher at $z \simeq 0.7$ than in the local Universe and could be responsible for the steep decrease in the SFR density from this epoch to the present one. 

Synthetic templates of star forming galaxies have demonstrated to be powerful tools to derive star formation related properties of galaxies, especially in the wavelength range UV-to-NIR (e.g. Leitherer \& Heckman 1999; Bruzual \& Charlot 2003; Fioc \& Rocca-Volmerange 1997).
Furthermore, many efforts have been devoted to model the properties of the dust emission in star forming galaxies (e.g. Dale et al. 2001; Chary \& Elbaz 2001; Rowan-Robinson et al. 2004). 
There have also been several attempts to reproduce the whole SEDs of star forming galaxies by using different methods: Silva et al. (1998) and Piovan et al. (2006) take into account the effects of the interstellar medium by considering a two phase model with a diffuse component and a dense one corresponding to molecular clouds. 
Devriendt et al. (1999) synthesize the stellar component and use a phenomenological fit to compute the extinction curve assuming a simple geometry for the dust, which yields the fraction of the stellar light reprocessed by the dust.
Popescu et al. (2000) solve the radiation transfer problem for a finite disk and a bulge by using a consistent treatment of grain heating and emission.
Finally, Dopita et al. (2005) perform a detailed modeling of the PAH emission features and of the reprocessing by the dust grains over a range of sizes.

In this paper we derive properties of a sample of UV selected galaxies by fitting multi-band data to a set of GRASIL (Silva et al. 1998) templates in a statistical way, similar to what has been done for other samples of galaxies by using UV-to-optical templates (e.g. Kauffmann et al. 2003) or UV-to-optical templates plus a total IR luminosity (Burgarella et al. 2005) built from other population synthesis codes. 
We obtain mean star formation related properties at different redshift bins and explore the likely evolution of these properties with redshift. 
The paper is organized as follows: In \S~2 the multi-band dataset is described. The main properties of the synthetic models are explained in \S~3. 
The fitting method and the determination of the parameters are described in \S~4 and 5 respectively. \S~6 contains a discussion on the main results obtained, and a brief summary is presented in \S~7. 
Throughout this paper we will use the following cosmological parameters: $H_{0} = 72$~km~s$^{-1}$~Mpc$^{-1}$, $\Omega_{M} = 0.3$ and $\Omega_{\lambda} = 0.7$.

\section{The dataset}

\subsection{Sample selection}

Our dataset contains multi-wavelength (UV to FIR) photometric data of galaxies in the ELAIS fields.
The UV data were taken by GALEX (Martin et al. 2005a) and consist of 19 circular frames (0.6~deg radius) in the ELAISN1 and ELAISN2 fields, observed through two filters with the bandpass centered at 1530\AA\ (FUV) and 2315\AA\ (NUV). 
In this paper we use only  NUV photometry, since FUV photometry is available only for a few of these frames. 
The final frames were the result of the co-add of different observations with typical total exposure times of $\approx 15000$~s. The NUV photometry used in this paper corresponds to the GALEX Second Data Release (available at http://galex.stsci.edu/GR2/). Table~\ref{galex_frames} shows the coordinates and exposure times of the GALEX frames used in this work.

The optical to FIR data correspond to the Spitzer Wide-area InfraRed Extragalactic (SWIRE) survey and consist of 12-band photometry: $U,g',r',i',Z$ taken at INT2.5m with the WFC; 3.6$\mu$m,4.5$\mu$m,5.8$\mu$m,8.0$\mu$m from Spitzer/IRAC; and 24$\mu$m,70$\mu$m,160$\mu$m from Spitzer/MIPS. In addition, photometric redshifts are provided for all the galaxies on the list. Throughout this paper, the merged catalog (version June 2005, Rowan-Robinson, private communication) was used. Details on the photometry of the SWIRE sources and on the quality of the photometric redshifts are given in Rowan-Robinson et al. (2005, hereafter RR05).

This work intends to study the properties of a UV selected sample of galaxies so we focused on the GALEX frames and selected all the sources with $m_{NUV} \leq 22.5$~mag. 
Sources located out of the inner 0.5~degree radius from the center of the GALEX frames were rejected since going to larger distances from the center can introduce spurious detections and ghosts that often appear close to the border. 
The angular resolution of the GALEX frames (FWHM $\approx 5$'') does not allow a proper star/galaxy separation, thus a cross-correlation between the UV and SWIRE catalogs was performed. 
All the GALEX sources for which a single SWIRE source (classified galaxy in the $g'$ and $r'$ bands) is present in a circle of a 2'' radius were included in our sample. 
In order to avoid confused detections due to the different angular resolution of the GALEX and (SWIRE) optical frames, those UV sources for which more than one optical detection was present within a circle of radius 6'' were excluded from the final sample. 
In this way, a magnitude limited UV selected sample of galaxies free of confusion effects was built.

\begin{figure}
\epsscale{1.0}
\plotone{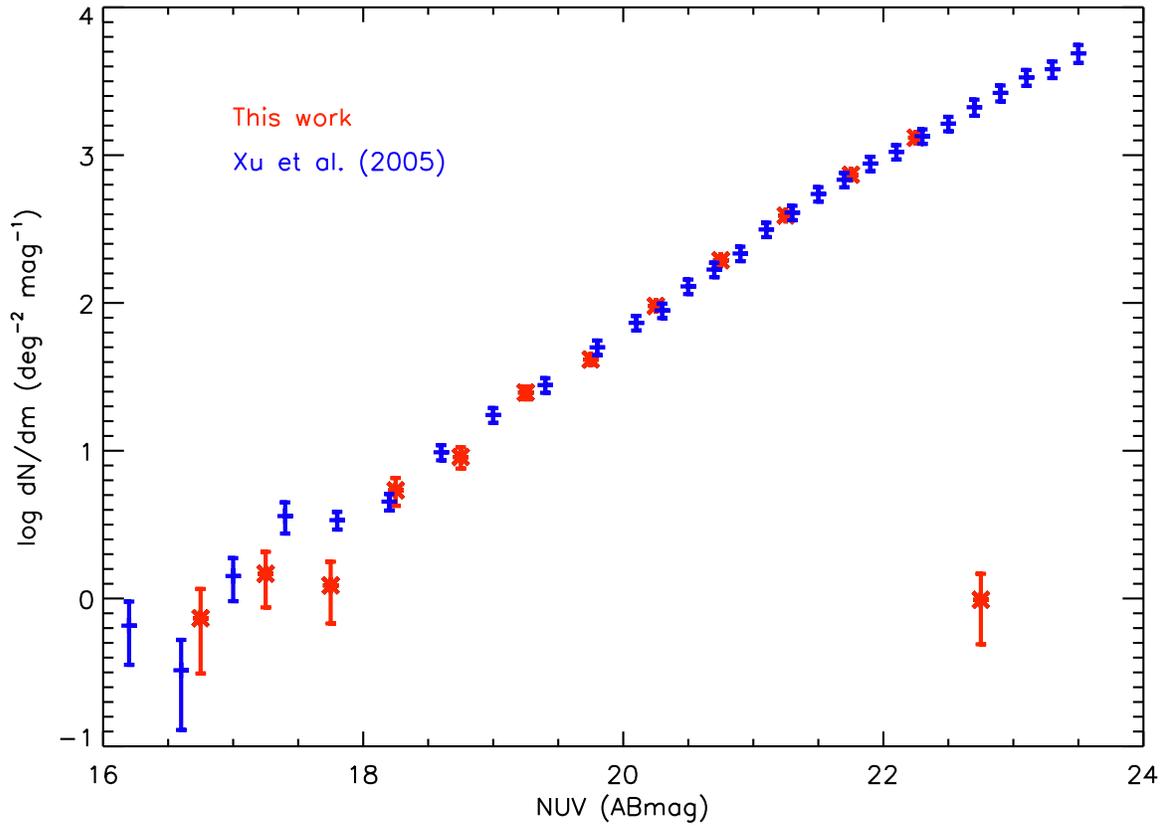}
\caption{NUV counts derived from our UV selected sample after matching with the optical-MIR SWIRE catalog (red). NUV counts derived by Xu et al. (2005) are also shown (blue). The histogram with the number of galaxies detected for each magnitude bin is also shown. \label{counts}}
\end{figure}

Both, the 6'' isolation condition previously mentioned and the cross-correlation with the SWIRE catalog may prevent a pure UV selection of our sample due to two reasons: (a) We may lose closely interacting galaxies and mergers which are either not resolved in the NUV frames (but they do in the SWIRE catalog) or are so close that they break the 6'' isolation condition; and (b) The SWIRE catalog contains only sources which were detected at both $r'$ and 3.6$\mu$m, which means that our final sample could lack blue galaxies included in our original NUV list but not detected either at $r'$ or at 3.6$\mu$m. 
However, the completeness limits at 3.6$\mu$m (3.7$\mu$Jy) and $r'$ (23~Vega mag) are much deeper than our NUV survey so including only SWIRE sources classified as galaxies does not introduce any bias in the sample (see Xu et al. 2007 for a discussion on this point). 
Also shown in Figure~\ref{counts} are the NUV counts derived from our sample compared to those of Xu et al. (2005). As can be seen, both sets of counts are consistent within the errors. Thus, we conclude that the selection procedure does not include significant biases and for practical reasons, our UV selected sample can be considered complete down to $m_{NUV} = 22.5$~ABmag.

\begin{figure}
\epsscale{1.00}
\plotone{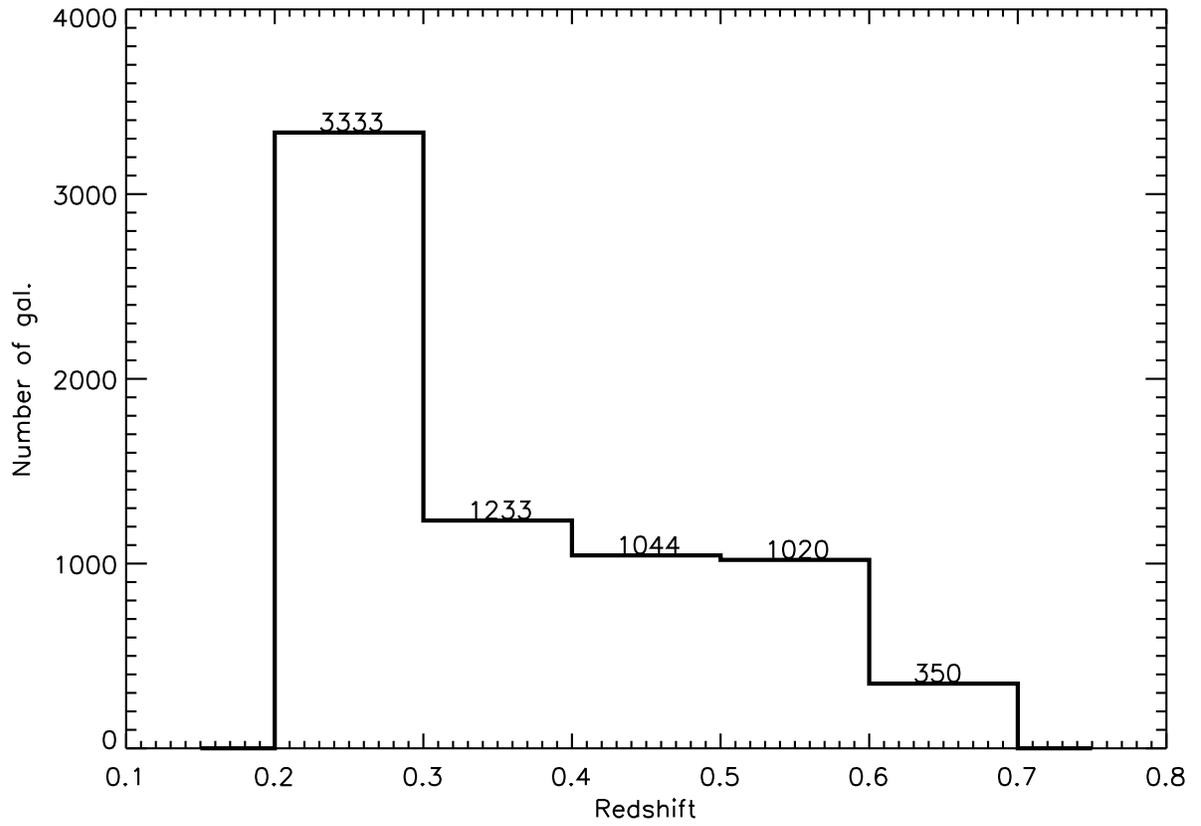}
\caption{Redshift distribution of the sample galaxies. The numbers on top of the histogram correspond to the number of galaxies in each redshift bin.\label{histo_red}}
\end{figure}

In the end we selected 6980 galaxies with (photometric) redshifts in the interval $0.2 \leq z \leq 0.7$. 
There are several reasons for this redshift restriction: firstly, the relative uncertainty of the photometric redshifts increases as they approach 0 (see RR05 for details), which will surely induce large uncertainties in the derived parameters dependent on redshift; secondly, Buat et al. (2006) have already analyzed the properties of a UV selected sample of galaxies at $z \approx 0$; finally, going to higher values of redshift would increase the number of non detections, mainly at IR wavelengths, and this fact would hamper our analysis. 
Figure~\ref{histo_red} shows the redshift distribution of our sample galaxies and
Table~\ref{comple} the fractions of detected galaxies at all the available bands.

\subsection{AGN contamination}

\begin{figure}
\epsscale{1.00}
\plotone{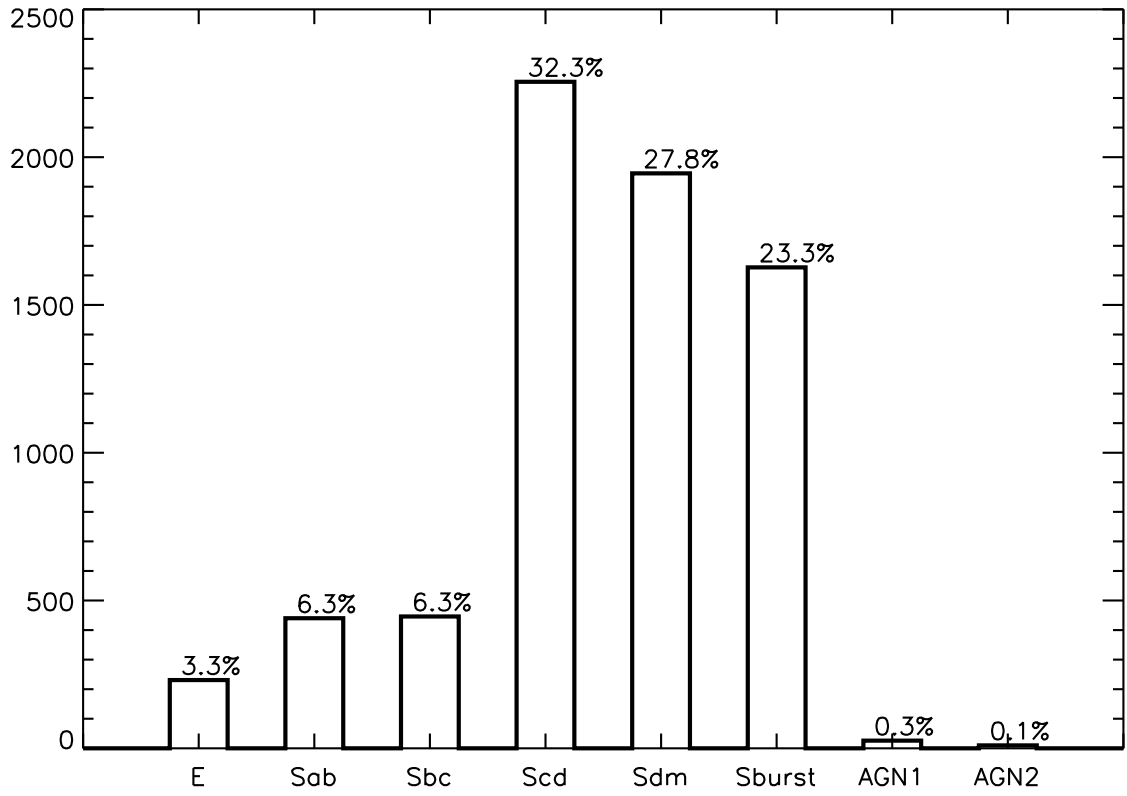}
\caption{Histogram of the best fit optical templates found by RR05 for our sample galaxies. The fractions of galaxies corresponding to each bin are indicated. \label{tem_opt}}
\end{figure}

Although a UV selected sample of galaxies is expected to be dominated by star forming galaxies, we cannot neglect the possibility of some contamination due to the presence of AGNs. In order to estimate the fraction of AGNs, we use the classification of RR05. 
In that paper the authors fitted the data using a set of 8 empirical optical-MIR templates\footnote{No UV fluxes were used in RR05.} corresponding to E, Sab, Sbc, Scd, Sdm, Sburst, AGN1 and AGN2. 
Figure~\ref{tem_opt} shows the distribution of their best fit templates for our 6980 galaxies. 
As can be seen, only about 0.4\% of the galaxies were classified as AGN by RR05. In addition, less than 4\% of the galaxies were classified as E, the rest of them being classified as spirals or starbursts. 
This result supports the fact that from the optical point of view our sample is dominated by star forming galaxies, with very low contamination from AGNs and ellipticals.

\section{Synthetic SEDs}

A set of synthetic templates of SEDs were computed with the code GRASIL\footnote{Details about the use of the code can be found at the web site http://web.pd.astro.it/granato/grasil/grasil.html. A web interface to use GRASIL can be found at http://web.pd.astro.it/galsynth/index.php.}. 
This code treats with particular care the effects of a dusty interstellar medium for different geometries of the dust and stars and is coupled with a chemical evolution code (CHE\_EVO) that takes into account in a consistent way the chemical evolution of the galaxy and the star formation history. 
This code describes one-zone open models depending only on time, including the infall of primordial gas. A more detailed description of this code can be found in Silva et al. (1998).
Although GRASIL allows the variation of a large set of free parameters (see Silva et al. 1998 for a detailed description of the code), to build our grid of templates only a small subset of parameters were varied (Silva, private communication).
The values spanned by the free parameters were chosen so that they cover the values proposed by Silva et al. (1998) to reproduce the SEDs of star forming galaxies of different Hubble types in the local Universe (see Silva et al. 1998 for details on fits to spiral and starburst galaxies). 

\begin{itemize}

\item The SFR was assumed to follow a Schmidt-type law of the form
\begin{equation}
SFR(t) = \nu M_{g}(t)^{k}
\end{equation}
where $M_{g}(t)$ is the gas mass at any time, $k = 1$, and $\nu$ is a free parameter varying from $\approx 2.56 \times 10^{-4}$ to $\approx 4$. 

\item Gas is continuously infalling at a rate proportional to $\exp(-t/\tau_{inf})$, where $\tau_{inf}$ is a free parameter taking values from 0.1 to 21.6~Gyr. These values extend above and below the typical values found for the Milky Way by Boissier \& Prantzos (1999): $\tau = 1$~Gyr at $R = 1$~kpc, and $\tau = 10$~Gyr at $R = 18$~kpc. GRASIL does not allow us to vary infall rates with the distance to the center of the galaxy, so single values are adopted for each model.

\item GRASIL also allows the addition at any time of an analytical extra star formation law (simulating a burst of star formation) combined with a Schmidt law. 
We chose an exponentially declining functional form with declining timescale equal to 50~Myr. 
For comparison, Kauffmann et al. (2003) and Salim et al. (2005) use bursts of constant intensity between 30 and 300~Myr after the beginning of the burst.
Four intensities of the burst were allowed in our work, corresponding to 0.33\%, 1\%, 3\% and 9\% of the total mass infall. 

\item GRASIL assumes two components of the interstellar medium (ISM): the molecular clouds -- containing the young stars and a dense ISM --, and the diffuse gas (cirrus) -- containing stars that escaped from the molecular clouds (free stars) and a diffuse ISM --. The typical time required to escape the molecular clouds, $t_{esc}$, is also a free parameter that we let vary from 1 to $4 \times 10^{7}$~yr, a value consistent with the one used by Charlot \& Fall (2000) and Panuzzo et al. (2007). An escape time of $2 \times 10^{7}$~yr is able to reproduce 
the SEDs of some local spiral galaxies (Silva et al. 1998).

\item The optical depth of the molecular clouds, $\tau_{MC} \propto m_{MC}/r^{2}_{MC}$, where $m_{MC}$ and $r_{MC}$ are the mass and radii of the molecular clouds respectively, was taken as another free parameter in our set of templates. We fixed $r_{MC} = 14$~pc and let $M_{MC}$ vary from $3 \times 10^{5}$ to $3 \times 10^{6}$~M$_{\odot}$. These values are close to the ones measured by Rosolowsky et al. (2007) for Giant Molecular Clouds in M33, although their masses are slightly lower and their radii slightly larger than the ones proposed here. On the other hand, our values are close to the ones used in Silva et al. (1998) to reproduce a sample of local spirals.

\item GRASIL allows us to stop the evolution of the galaxy at any time, so that galaxies of different ages can be simulated. We assumed that all galaxies are coeval with an age of $\approx 13$~Gyr. For the redshift interval selected in this work, and given that SEDs are almost insensitive to small age differences, we produced models of 8 and 12~Gyr.

\item The inclination angle of the galaxies may play a role in the observed SED of a galaxy, since the more inclined a galaxy is, the more attenuated it appears at UV and optical wavelengths\footnote{The IR emission remains almost unchanged since it is isotropic}. This may result in an incorrect derivation of the dust attenuation and star formation related parameters. Although we lack available information on the inclination angles for our galaxies, we generated models corresponding to inclination angles of 15$^\circ$, 45$^\circ$ and 75$^\circ$. 

\end{itemize}

In all our models, stars and gas are distributed following exponential disks with scale length 4~kpc and scale height 0.4~kpc. 
These values are of the order of the ones reported for large samples of spiral galaxies (de Jong et al. 1996; Yoachim \& Dalcanton 2006) although other authors report slightly higher values for the scale height (Barteldrees \& Dettmar 1994).
The assumed fraction of molecular gas is 50\%, which is rather high but still reasonable for star forming galaxies (Boselli et al. 2002).
Finally, the dust to gas ratio is assumed to be proportional to the metallicity and normalized to the Milky Way value at solar metallicity.

Our database is composed of a total of 5127 templates.
Table~\ref{templates} lists the free parameters of the templates used in this paper.

The evaluation of the total IR luminosity between 8 and 1000$\mu$m ($L_{IR}$) is a potential source of uncertainty, mainly when almost no detections at FIR wavelengths and detections at 24$\mu$m for only about half of the sources are available, as is the case for our dataset.
For this reason, we prefer to be consistent with previous observational results and we restrict
our models to those which satisfy the following relations: 
\begin{equation}
\label{eq_ir8}
L_{IR} = 10^{0.28} \times L_{8}^{1.060}
\end{equation}
\begin{equation}
\label{eq_ir25}
L_{IR} = 10^{2.01} \times L_{25}^{0.878}
\end{equation}
\begin{equation}
\label{eq_ir60}
L_{IR} = 10^{0.40} \times L_{60}
\end{equation}
where $L_{8}$, $L_{25}$ and $L_{60}$ are the monochromatic luminosities at 8, 25 and 60$\mu$m respectively. These three observational constraints have been reported by Caputi et al. (2007) and Takeuchi et al. (2005b) and hold over several orders of magnitude. 
In order to account for the dispersions found around these relations, we allow models to deviate by a factor of 3 above and below these average relations. 
We note that whereas two of these relations are found for local galaxies (Takeuchi et al. 2005b), the one by Caputi et al. (2007) holds for galaxies at $0 < z < 2$. 
Thus, by applying these local relations to our models we assume that the IR SEDs are suitable to estimate the total IR luminosities of higher redshift galaxies, as seems to be the case (Egami et al. 2004; Le Floch et al. 2005; Marcillac et al. 2006).

In order to show how our models reproduce the SEDs of the sample galaxies from the UV up to the IR, in Figures~\ref{gr_ri_z0p25} to \ref{nuvg_gr_z0p25} several observed color-color plots for the galaxies and for the templates at different redshifts are presented. 
The observed colors involving only optical bands (figure~\ref{gr_ri_z0p25}) are properly reproduced by the models at any redshift. The fraction of galaxies located at distances larger than 0.2~mag from any template is lower than 3\% at all redshifts. Concerning the observed colors involving optical and IR data (figure~\ref{gr_r3p6_z0p25}) the situation is quite similar, with less than 4\% of the galaxies more distant than 0.2~mag to any template. 
However, the situation is less favourable for the observed colors involving the NUV fluxes, as is shown in figure~\ref{nuvg_gr_z0p25}. In this case, the fraction of galaxies showing $(NUV - g')$ colors not reproduced by the models to an accuracy of 0.2~mag increases from 5\% for $z \in [0.2,0.3]$ to 25\% for $z \in [0.6,0.7]$. The reason for this discrepancy between data and models is not clear. A careful revision of the most extreme cases yields NUV fluxes much larger (compared to their optical fluxes) than those expected for star forming galaxies. The median NUV fluxes of the discrepant galaxies are similar to that of the total sample, so photometric uncertainties can be discarded as the only reason for this effect. 
In any case, the total number of discrepant galaxies is less than 10\%, therefore most of the galaxies can be properly fitted by the models. 

\section{SED fitting procedure}

For each galaxy of our sample the reduced $\chi^{2}$ corresponding to all the available templates satisfying Eqs.~\ref{eq_ir8}, \ref{eq_ir25} and \ref{eq_ir60} is computed. In this way the scaling factors for all the templates are obtained. 
Many of our galaxies have no available IR fluxes, so the upper limits in the derivation of $\chi^{2}$ are included.

Following the usual notation, $\chi^{2}$ is defined as:
\begin{equation}
\chi^{2} = \frac{1}{n - 1}\sum_{i} \chi^{2}_{i}
\end{equation}
where $n$ is the number of bands at which the galaxy was detected or for which the model flux exceeds the upper limit flux at any wavelength, and
\begin{eqnarray}
\chi^{2}_{i} & = & 
\left\{ 
\begin{array}{lll}
(f_{\lambda,i} - j_{\lambda,i})^{2}/(\Delta f_{\lambda,i})^{2}& , & \mbox{if detection at } \lambda_{i} \\
(f^{u}_{\lambda,i} - j_{\lambda,i})^{2}/(\Delta f^{u}_{\lambda,i})^{2}& , & \mbox{if non detection at } \lambda_{i} \mbox{ and } f^{u}_{\lambda,i} \leq j_{\lambda,i} \\
               0 & , & \mbox{if non detection at } \lambda_{i} \mbox{ and } f^{u}_{\lambda,i} > j_{\lambda,i} \\
\end{array}
\right.
\nonumber \\
\end{eqnarray}
where $f_{\lambda,i}$ is the detected flux at $\lambda_{i}$, $\Delta f_{\lambda,i}$ is the flux uncertainty at $\lambda_{i}$, $f^{u}_{\lambda,i}$ is the upper limit at $\lambda_{i}$, $\Delta f^{u}_{\lambda,i}$ is the uncertainty associated to $f^{u}_{\lambda,i}$, and $j_{\lambda,i}$ is the model flux at $\lambda_{i}$.

\begin{figure}
\epsscale{1.00}
\plotone{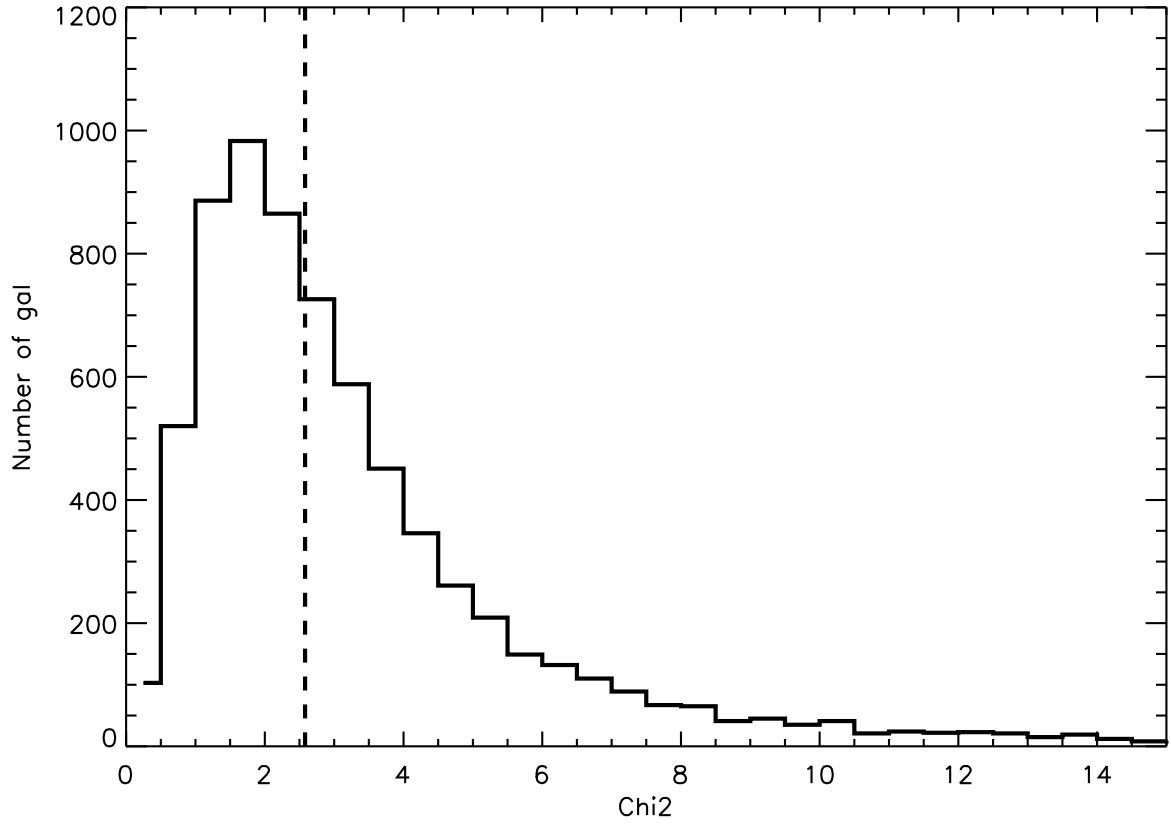}
\caption{Reduced $\chi^{2}$ distribution of the sample galaxies. The vertical line indicates the median value $\chi^{2} = 2.58$.\label{histo_chi2}}
\end{figure}

For each galaxy, the associated value of the reduced $\chi^{2}$ is the minimum among all the templates.
Figure~\ref{histo_chi2} shows the histogram of the reduced $\chi^{2}$, the median value being 2.58. 
In order to illustrate the quality of the fits for different values of $\chi^{2}$, a selection of them are shown in Figure~\ref{plot}. 
It can be seen that the quality of fits with $\chi^2 \leq 4$ is reasonably good.
For our analysis, we prefer to keep only galaxies with reasonably low values of $\chi^2$ since otherwise, the uncertainties in the derived parameters will be large. For this reason, we exclude all the galaxies for which $\chi^{2} > 4$, and we end up with a sample of 5133 galaxies.
We have verified that discarding these galaxies does not introduce any bias in our sample: the fraction of galaxies detected at 24$\mu$m, the redshift distribution of galaxies and the average values of rest frame $L_{FUV}$ are similar for the complete and the restricted (good quality) samples.
In addition, by imposing a limiting $\chi^{2}$ we also reduce the fraction of galaxies with discrepant $(NUV - g')$ color which is finally of 1\%, 1.5\%, 7\%, 5\% and 11\% for $z \in [0.2,0.3], [0.3,0.4], [0.4,0.5], [0.5,0.6]$ and $[0.6,0.7]$ respectively.

\section{Determination of parameters and associated uncertainties}

For each template, GRASIL provides a complete SED (UV to FIR), a list of parameters related to the evolution of the galaxy such as the SFR and mass assembly as a function of time. For our purposes, we derive the luminosity\footnote{Throughout this paper, monochromatic luminosities are defined as $\lambda F_{\lambda}$, where $\lambda$ is the reference wavelength and $F_{\lambda}$ is the flux density at this wavelength.} at any wavelength from the output SED. The total $L_{IR}$ is the luminosity emitted between 8 and 1000$\mu$m. 
The SFR and stellar mass (M*) are directly given by GRASIL. In this paper we will consider the SFR averaged over the last $10^{8}$~yr, since this is the typical timescale of star formation to which the UV and IR fluxes are sensitive (e.g. Iglesias-P\'{a}ramo et al. 2006). 

The physical parameters associated to a given galaxy are the median values of the likelihood distribution resulting from applying the bayesian statistics to the whole set of templates, as described in detail in Kauffmann et al. (2003). 
The confidence intervals are defined as those containing 95\% of the likelihood distribution. 

Figures~\ref{confi_lfuv} to \ref{confi_lir} show the distribution of the confidence intervals corresponding to $L_{FUV}$, $M*$ and $L_{IR}$. As can be seen, the $L_{FUV}$ and $M*$ distributions are clearly unimodal. This is not unexpected since the contribution of the stars to the SED mostly ranges from the UV to the NIR, which is reasonably well covered in our dataset. The median values of the confidence intervals are 0.34 and 0.52~dex respectively for $L_{FUV}$ and $M*$. 

On the contrary, the $L_{IR}$ distribution shows a bimodal shape, 
with a median value of the confidence interval of 0.73~dex.
This different behavior is due to the paucity of the coverage at FIR wavelengths of our dataset. In fact, the fraction of galaxies detected at 24/70/160$\mu$m (the wavelengths contained in the interval along which $L_{IR}$ is computed) is about 47/3/1\%\footnote{These fractions correspond to the good quality sample after excluding the galaxies with $\chi^{2} \geq 4$.}. This means that for about 50\% of the sample, $L_{IR}$ is extrapolated from the GRASIL SEDs which are constrained mostly at UV-optical-MIR wavelengths. 

In order to evaluate the quality of the values of $L_{IR}$ derived from the GRASIL models, we compared our results with the 
models of Chary \& Elbaz (2001). 
These models provide $L_{IR}$ whenever the redshift of the galaxy and the flux at 24$\mu$m are known. 
For this reason, a direct comparison is only possible for the subsample of galaxies detected at 24$\mu$m. 
The comparison is shown in Figure~\ref{lir_lirelbaz}. 
Our determination of $L_{IR}$ is larger than that of Chary \& Elbaz (2001) by 25\% on average, with a dispersion of 0.13~dex. If we restrict ourselves to those galaxies detected at 70$\mu$m, our estimation is still 4\% larger with a dispersion of 0.12~dex.

\section{Results and discussion}

\subsection{Dust attenuation versus total luminosity and stellar mass}

$L_{bol}$ (defined as $L_{FUV} + L_{IR}$) is a good tracer of the total luminosity emitted by young stars since it contains both the contribution of the light directly emitted by young stars plus the IR emission of the dust, which absorbs and scatters part of this light (e.g. Martin et al. 2005b; Buat et al. 2006). 
The energetic budget between the IR and UV emission (namely the $L_{IR}/L_{FUV}$ ratio) is known to be a good tracer of the dust attenuation at UV wavelengths (Xu \& Buat 1995; Meurer et al. 1999; Gordon et al. 2000).
It has already been reported in the literature that galaxies with large SFRs also tend to show a larger dust attenuation (Wang \& Heckman 1996; Buat \& Burgarella 1998; Hopkins et al. 2001). 
Similar relations have also been found for galaxies at higher redshifts (Burgarella et al. 2006; Reddy et al. 2006) selected at UV rest frame. 
With our sample we have the opportunity to study the evolution with redshift of this relation for UV selected galaxies from $z \approx 0.7$ to $z \approx 0.2$.

Figure~\ref{lbol_lirlfuv} shows $L_{FUV}/L_{IR}$ as a function of $L_{bol}$. 
We have split our sample into five redshift bins: $[0.2,0.3]$, $[0.3,0.4]$, $[0.4,0.5]$, $[0.5,0.6]$ and $[0.6,0.7]$. 
The top-left panel also shows the average relation of Buat et al. (2006) for UV selected local galaxies with the same limiting magnitude as our subsample with $z \in [0.2,0.3]$.
As can be seen, for $L_{bol} \geq 10^{10}$~L$_{\odot}$ the agreement is quite remarkable with the data at $z \in [0.2,0.3]$. 
For the sake of simplicity and homogeneity we constraint our study of the evolution of the $L_{IR}/L_{FUV}$ ratio to our subsamples at different redshift bins.

The  comparison between the different redshift bins is  difficult because the observational limit ($m_{NUV} \leq 22.5$~ABmag) translates to different luminosity limits, thus we can only samples of galaxies in regions of the plot which are free of observational biases can be compared. 
The observational limits for each redshift bin were estimated by assuming that the limiting NUV magnitude translates into $L_{FUV}$ rest frame and obtained limiting values of $L_{FUV} = 10^{9}, 2.25 \times 10^{9}, 4.95 \times 10^{9}, 8.46 \times 10^{9}$ and $1.32 \times 10^{10}~L_{\odot}$ for $z = 0.2, 0.3, 0.4, 0.5$ and 0.6 respectively.
The limit at $z = 0.6$ restricts the comparison to a very reduced number of galaxies at lower redshifts in Figure~\ref{lbol_lirlfuv}. For this reason we perform our comparison within regions (a) and (b) of the figure, which are free of observational biases for galaxies with $z \in [0.2,0.6]$. These two regions correspond to $L_{bol} \in [2.5 \times 10^{10},10^{11}]$ and $[10^{11},4 \times 10^{11}]$~$L_{\odot}$.

We rely on the fact that galaxies should be equally distributed in any region of the diagram free of observational biases, if the dust attenuation does not evolve with redshift for a given total luminosity. 
Thus, a detailed study of the distributions of galaxies in regions (a) and (b) is performed at different redshift bins. 
The results of a Kolmogorov-Smirnov test comparing the populations of both regions for different redshifts are shown in Table~\ref{ks}. 
For region (a) only galaxies with $z \in [0.3,0.4]$ show distributions consistent with those at $z \in [0.2,0.3]$ at the 95\% confidence level. For the rest of the redshift bins, the distributions of galaxies are significantly different to each other.  
On the other hand, the distributions of galaxies in region (b), are consistent with being drawn from the same population at the 95\% confidence level, so no evolution with redshift is reported for them. Figure~\ref{histo_lbol}(a) shows that the histograms of $L_{IR}/L_{FUV}$ for region (a) are shifted towards higher values as redshift decreases, which suggests that low total luminosity galaxies -- region (a) -- with low values of $L_{IR}/L_{FUV}$ become scarce at low values of $z$. This decreasing presence of galaxies with low $L_{IR}/L_{FUV}$ as redshift decreases is not seen for high total luminosity galaxies -- region (b) -- as the Kolmogorov-Smirnov test states and Figure~\ref{histo_lbol}(b) suggests.

The evolution of the dust attenuation with the stellar mass is also studied. Figure~\ref{mstar_lirlfuv} shows $M*$ vs. $L_{IR}/L_{FUV}$ at different redshifts is also studied. 
Again we are only limited by our UV selection because the rest of the relevant parameters ($M*$, $L_{IR}$) are derived from the templates after fitting to the observational data. The only limit in Figure~\ref{mstar_lirlfuv} at a given redshift is a minimum $M*$ associated to the minimum $L_{FUV}$ imposed by the UV selection. Thus, the distribution of stellar masses represented in this figure corresponds to that of a UV selected sample and does not intend to be complete down to any value of $M*$.

In Figure~\ref{mstar_lirlfuv_mean} the volume weighted average
$M*$ vs. $L_{IR}/L_{FUV}$ at the different redshifts is shown. 
The average and standard deviation (corresponding to the error bars in the figure) values have been estimated as in Buat et al. (2006):

\begin{equation}
\left< L_{IR}/L_{FUV} \right> = \frac{\Sigma_{i} \omega_{i} (L_{IR}/L_{FUV})_{i}}{\Sigma_{i} \omega_{i}}
\end{equation}
and
\begin{equation}
\sigma^{2}(L_{IR}/L_{FUV}) = \frac{\Sigma_{i} \omega_{i} ((L_{IR}/L_{FUV})_{i} - \left< L_{IR}/L_{FUV} \right>)^{2}}{\Sigma_{i} \omega_{i}}
\end{equation}
where $\omega_{i} = 1/V_{max}$ for galaxy $i$.

It can be seen that $L_{IR}/L_{FUV}$ increases with stellar mass at all redshifts 
for galaxies with $M* \leq 10^{11}$~M$_{\odot}$.
A galaxy with $M* \approx 2 \times 10^{10}$~M$_{\odot}$ 
shows $L_{IR}/L_{FUV}$ a factor of $\simeq 3$ higher at $z = 0.65$ than at $z = 0.25$.
This difference is lower for a galaxy with $M* \approx 7 \times 10^{10}$~M$_{\odot}$, a factor of $\simeq 2$.
For galaxies with $M* \geq 10^{11}$~M$_{\odot}$ the trend seems to flatten and $L_{IR}/L_{FUV}$ reaches an almost constant value with redshift.
The plot also shows that for galaxies more massive than $\log M* = 11.5$ the evolutionary trend of $L_{IR}/L_{FUV}$ seems to invert in the sense that it decreases with redshift. Whether this effect is real or not is still to be confirmed since the statistics are scarce for this last mass bin.

A tentative explanation of this evolution of $L_{IR}/L_{FUV}$ with the stellar mass could be associated with the total dust content of star forming galaxies, which is linked to the metallicity. A mass--metallicity relation has been reported at different redshifts in the sense that more massive galaxies are also more metal rich (Zaritsky et al. 1994, Tremonti et al. 2004). We also find that at a fixed stellar mass, the dust attenuation increases as redshift decreases. This result is consistent with the findings of Savaglio et al. (2005) and Erb et al. (2006) that for a given stellar mass, galaxies were less metallic at high redshift than they are at present time. This result suggests a mass dependent rate of chemical enrichment for galaxies that can be explained by invoking a mass dependent star formation history. In fact, N-body simulations by Brooks et al. (2007) claim that low star formation efficiencies are required for low mass galaxies in order to explain the mass--metallicity relation. 
We can estimate the star formation efficiency of our galaxies through the fitted parameters of GRASIL. The $\nu$ parameter is the main driver of the SFR  when we assume a Schmidt-type law (Eq.~1). Low values of $\nu$ allow a fairly flat SFH until the present epoch. 
On the contrary, high values of $\nu$ imply a very intense star formation activity in the first epochs of a galaxy followed by a steep decrease of the SFR due to a quick exhaustion of the gas reservoir. 
This decrease in the SFR of massive galaxies has already been reported by Bell et al. (2005), Juneau et al. (2006) and Pannella et al. (2007) for different samples of galaxies from $z \approx 2$ to present time.
Figure~\ref{mstar_nu} shows the relation between M* and $\nu$ for our galaxies. It is apparent that low mass galaxies show lower values of $\nu$ while the opposite holds for high mass galaxies. 
With this downsizing picture in mind we can explain the differential evolution of dust attenuation as partially linked to the SFH of galaxies: high mass galaxies experience a first epoch of intense star formation activity, thus they will efficiently increase their metal (and dust) content. As the SFR begins to decline up to the present epoch the metal enrichment becomes slower as does the dust production. 
Conversely, low mass galaxies show almost constant star formation efficiencies, which  translate into moderate but continuously increasing chemical enrichment and dust production. 
To illustrate this hypothesis the volume weighted metallicity evolution for different bins of M* is shown in Figure~\ref{zeta_mass}. Metallicities have been computed by the chemical evolutionary code CHE\_EVO coupled to GRASIL in a consistent way with the rest of the parameters, and we remind the reader that in our models the dust to gas fraction was assumed to be proportional to metallicity. 
As can be seen, whereas for galaxies with $\log M* \in [11.5,12]$ the metallicity increases by a factor of $\simeq 1.6$ from $z = 0.65$ to $z = 0.25$, for galaxies with $\log M* \in [10,10.5]$ the increasing factor is $\simeq 8$. 
Thus the evolution of $L_{IR}/L_{FUV}$ observed in figure~\ref{mstar_lirlfuv_mean} could be understood in terms of an evolution of metallicity.

\subsection{The specific SFR}

The specific star formation rate (SSFR) is defined as $SSFR = SFR/M*$, that is, the ratio of the current SFR to the total stellar mass produced throughout the galaxy's lifetime. 
It is related to the birthrate parameter $b$ (defined as $SFR/\left< SFR \right>$, where $\left< SFR \right>$ is the averaged SFR throughout the galaxy's lifetime), but it does not depend on the timescale required to build all the stellar mass. If we assume that the SFR of the galaxy will remain almost constant for a long period of time, the SSFR is equivalent to the inverse of the time required by a galaxy to double its stellar mass.

Figure~\ref{mstar_ssfr} shows the SSFR as a function of the stellar mass. 
The observational limits are indicated with dashed lines. The limits correspond to the minimum stellar mass and SFR allowed for each $L_{FUV}$ of our sample.
The horizontal lines in each panel indicate the SSFR required for a galaxy to double its stellar mass from each epoch to present time assuming constant SFR.
It is found that less massive galaxies tend to have a larger SSFR, consistent with the scenario already reported by several authors (Cowie et al. 1996; Brinchmann \& Ellis 2000; Boselli et al. 2001; Brinchmann et al. 2004; Heavens et al. 2004; Feulner et al. 2005; Bell et al. 2005; Buat et al. 2006) for samples of galaxies selected on the basis of different wavelengths and at different redshifts.
For $z \in [0.2,0.3]$ most of the galaxies are below the doubling-mass line, but this trend changes at higher redshifts and at $z \in [0.6,0.7]$ most of the galaxies are well above this line\footnote{Feulner et al. (2005) call the galaxies falling below and above this line ``quiescent'' and ``star-forming'' respectively.}. 
This result is clearer in Figure~\ref{mstar_ssfr_volumen} where the volume weighted average SSFR as a function of M* for the different redshift bins are shown. 

For a given M*, the SSFR  is found to decrease with  redshift. The slope of the variation of the SSFR with M* as well as the rate of decline of SSFR with $z$ for different  stellar masses are clues to understand the origin of the downsizing process.
We find a shallower variation of SSFR  versus M* (i.e. a larger SSFR for massive galaxies) as compared to other works  aimed at studying the evolution of the SSFR-M* variation and covering (at least) the same redshift range (e.g. Feulner et al. 2005, Bauer et al. 2005,  Zheng et al. 2007). From Figure~\ref{mstar_ssfr_volumen}, the SSFR is also found to decrease by a larger factor for massive galaxies than for low mass ones from $z = 0.7$ to $z = 0$. As an example, the SSFR lowers by a factor of $\simeq 8$ from $z \approx 0.65$ to $0.25$ for galaxies more massive than $10^{11}$~M$_{\odot}$, whereas this factor is of the order of 4 for galaxies with $M* \approx 10^{10}$~M$_{\odot}$.
This  result differs from the recent findings of  Zheng et al. (2007) who conclude a similar decrease of SSFR  with redshift for all stellar masses from $z = 0.9$ to present time.

In this respect, the wavelength at which samples are selected may introduce some bias on the obtained results. 
To illustrate this point, Figure~\ref{secuencia} shows $M*$ vs. rest frame $(U - V)$ for our galaxies. The lines separating the red and blue sequences at the different redshifts (from Bell et al. 2007) are shown in separated panels. As can be seen, the majority of our galaxies belong to the blue sequence at all redshifts and almost no signs of any color bimodality are found (e.g. Strateva et al. 2001). This is a consequence of the UV selection applied to our sample. 
Contrary to that, the sample of Elbaz et al. (2007) selected at rest frame optical wavelengths shows a clear bimodality in the color magnitude plot. The same is found by Bell et al. (2007) for their sample selected at optical magnitudes. These authors also show that red galaxies show a steeper relation between $M*$ and SSFR than blue ones. 
From a general point of view, care should be taken when comparing star formation related properties from samples selected with different criteria since the results will depend on the fraction of red to blue galaxies of each sample. In particular, our results show that UV selected samples are extremely efficient at selecting blue galaxies. However, as has been shown in previous papers (Buat et al. 2005; Iglesias-P\'{a}ramo et al. 2006), a complete view of the evolution of star forming galaxies would require a sample of galaxies containing both the bright (UV) and dark (IR) sides of star formation.

\section{Conclusions}

In this paper a study has been performed on the properties of a sample of UV selected galaxies at intermediate redshifts ($0.2 \leq z \leq 0.7$). The properties were inferred from the fit to a grid of synthetic models constructed with the code GRASIL through a Bayesian analysis.
Most of the galaxies fit reasonably well to the models, which indicates that a Schmidt-type law combined with recent bursts of star formation properly reproduces the observed SEDs of galaxies at this redshift range. 

It was found that, whereas low luminosity galaxies with low attenuation were significantly more numerous at high redshifts, this result is not seen for high luminosity galaxies.
In the same way, dust attenuation is found to increase with decreasing redshift for galaxies with low $M*$, but this trend seems to disappear for galaxies with $\log M* \geq 11$.
This result is consistent with a mass dependent evolution of the dust to gas ratio, and thus of metallicity, which could be driven by a mass dependent star formation history of galaxies: 
the most massive galaxies have formed most of their stars by $z \simeq 0.7$ and thus their metallicity would not significantly change from $z = 0.7$ to $z = 0.2$, and likewise their dust content. The opposite is found to hold for low mass galaxies, which experience a larger increase in metallicity over the same period, producing the reported evolution in the dust attenuation.

The SSFR were found to decrease with increasing stellar mass, as has been known for many years (e.g. Cowie et al. 1996), and the trend holds at all redshifts explored in this work. 
Assuming constant SFRs, the fraction of galaxies able to double their stellar masses from their epoch to the present time decreases with redshift. At $z \geq 0.6$ most of the galaxies   with $M* \geq 2 \times 10^{10}~M$ are above the doubling--mass line whereas the opposite is true for such massive galaxies at $z \simeq 0.2$.
A shallower decrease of the SSFR with increasing stellar mass is found than reported in previously published studies in the same redshift range and based on visible-NIR selections.
This difference is likely to be due to the sample selections: whereas samples selected at optical/IR wavelengths include galaxies belonging to both the blue and red sequences, our UV selection is dominated by blue (star forming) galaxies.

\acknowledgments

We thank the anonymous referee for his/her interesting comments and suggestions that helped to improve the former version of this manuscript.
Thanks are also given to Steve Donegan for his careful reading and English revision of the text.
This paper has benefited from interesting discussions with G.L. Granato, L. Silva and P. Panuzzo. 
GALEX (Galaxy Evolution Explorer) is a NASA Small Explorer, launched in April 2003. We gratefully acknowledge NASA's support for construction, operation, and science analysis for the GALEX mission, developed in cooperation with the Centre National d'Etudes Spatiales of France and the Korean Ministry of Science and Technology. 
Support for this work, part of the Spitzer Space Telescope Legacy Science Program, was provided by NASA through an award issued by JPL under NASA contract 1407.
This article is based on observations made with the Isaac Newton Telescope, operated on the island of La Palma by the Isaac Newton Group at the Spanish Observatorio de El Roque de los Muchachos.
This work has been partially funded by the projects AYA 2004-08260-C03-02 of the Spanish PNAYA and TIC114 of the Junta de Andaluc\'{\i}a.
TTT has been supported by Program for Improvement of Research
Environment for Young Researchers from Special Coordination Funds
for Promoting Science and Technology (SCF) comissioned by the
Ministry of Education, Culture, Sports, Science and Technology
(MEXT) of Japan.

\newpage

\begin{table}

\caption{Main properties of the GALEX frames: (1) Id. of the GALEX frames; (2) Right Ascension of the center of the frame, in J2000.0 equinox; (3) Declination of the center of the frame, in J2000.0 equinox; (4) Total exposure time of the co-added NUV frame.}

\label{galex_frames}

\begin{tabular}{lccc}
\tableline
Name & R.A. & Dec. & $t_{exp}$ (NUV) \\
     & (J2000.0) & (J2000.0) & (s) \\
\tableline
ELAISN1\_00 & 243.40356 & 54.98425 & 14568.1 \\
ELAISN1\_01 & 241.96040 & 55.63836 & 15271.05 \\
ELAISN1\_02 & 241.72232 & 54.58142 & 14825.05 \\
ELAISN1\_03 & 245.14947 & 55.89300 & 13963.25 \\
ELAISN1\_04 & 243.27660 & 53.89144 & 14269.05 \\
ELAISN1\_05 & 243.15825 & 56.54596 & 14732.55 \\
ELAISN1\_06 & 241.50000 & 56.60900 & 14492.05 \\
ELAISN1\_07 & 244.70934 & 55.10306 & 14272.15 \\
ELAISN1\_08 & 241.00900 & 55.05406 & 13758.2 \\
ELAISN1\_09 & 241.60702 & 53.53426 & 13341.1 \\
ELAISN1\_10 & 240.40000 & 53.79400 & 13573.15 \\
ELAISN1\_11 & 242.82000 & 52.90000 & 14866.05 \\
ELAISN1\_12 & 244.11200 & 54.00000 & 15135.05 \\
ELAISN2\_00 & 249.21294 & 41.03876 & 27196 \\
ELAISN2\_01 & 247.82121 & 41.06719 & 14368.8 \\
ELAISN2\_03 & 250.53587 & 41.37288 & 14949.1 \\
ELAISN2\_04 & 250.15115 & 40.26699 & 15303.05 \\
ELAISN2\_05 & 248.55000 & 40.05000 & 13789.05 \\
ELAISN2\_08 & 249.66500 & 41.90100 & 14689 \\
\tableline
\end{tabular}

\end{table}

\newpage

\begin{table}

\caption{Detection rates at all bands for the sample galaxies.}
\label{comple}
\begin{tabular}{lcccccc}
\tableline
Band & NUV & U & g' & r' & i' & Z \\
\tableline
Det. rate (\%) & 100.0 & 90.0 & 100.0 & 100.0 & 100.0 & 98.1 \\
\tableline
\end{tabular}

\vspace{1cm}

\begin{tabular}{lccccccc}
\tableline
Band & 3.6$\mu$m & 4.5$\mu$m & 5.8$\mu$m & 8.0$\mu$m & 24$\mu$m & 70$\mu$m & 160$\mu$m \\
\tableline
Det. rate (\%) & 100.0 & 91.9 & 40.1 & 62.7 & 49.9 & 4.6 & 1.6 \\
\tableline
\end{tabular}

\end{table}

\newpage

\begin{table}

\caption{Free parameters of the GRASIL models.}

\label{templates}

\begin{tabular}{lc}
\tableline
SFR & SL, SL$+0.33$\%, SL$+1$\%, SL$+3$\%, SL$+9$\% \\
$t_{esc}$ (Gyr) & 0.01, 0.02, 0.04 \\
$M_{mol}$ (M$_{\odot}$) & $3 \times 10^{5}$, $10^{6}$, $3 \times 10^{6}$ \\
$\tau$ (Gyr) & 0.1, 0.6, 3.6, 21.6 \\
$\log \nu$ & $-3.6$, $-2.9$, $-2.2$, $-1.5$, $-0.8$, $-0.1$, $0.6$ \\
Age (Gyr) & 8, 12 \\
Incl. Angle & 15$^{\circ}$, 45$^{\circ}$, 75$^{\circ}$ \\
\tableline
\end{tabular}

\end{table}

\newpage

\begin{table}

\caption{Results of the Kolmogorov-Smirnov test for the distributions of $L_{IR}/L_{FUV}$ at different redshifts.}

\label{ks}

\begin{tabular}{llccc}
\tableline
Pop. $i$ & $N_{gal,i}$ & Pop. $j$ & $N_{gal,j}$ & $P_{i,j}$ \\
\tableline
\multicolumn{5}{c}{Region (a) $L_{bol} \in [2.5 \times 10^{10},10^{11}]$ } \\
\tableline
$z \in [0.2,0.3]$ & 100 & $z \in [0.3,0.4]$ & 119 & 0.54 \\
\multicolumn{2}{c}{...} & $z \in [0.4,0.5]$ & 292 & $2.19\times 10^{-6}$ \\
\multicolumn{2}{c}{...} & $z \in [0.5,0.6]$ & 334 & $1.48\times 10^{-20}$ \\
$z \in [0.3,0.4]$ & 119 & $z \in [0.4,0.5]$ & 292 & $5.87 \times 10^{-5}$ \\
\multicolumn{2}{c}{...} & $z \in [0.5,0.6]$ & 334 & $1.99\times 10^{-17}$ \\
$z \in [0.4,0.5]$ & 292 & $z \in [0.5,0.6]$ & 334 & $2.76 \times 10^{-9}$ \\
\tableline
\multicolumn{5}{c}{Region (b) $L_{bol} \in [10^{11},4 \times 10^{11}]$ } \\
\tableline
$z \in [0.2,0.3]$ & 70 & $z \in [0.3,0.4]$ & 88 & 0.64 \\
\multicolumn{2}{c}{...} & $z \in [0.4,0.5]$ & 279 & 0.67 \\
\multicolumn{2}{c}{...} & $z \in [0.5,0.6]$ & 396 & 0.19 \\
$z \in [0.3,0.4]$ & 88 & $z \in [0.4,0.5]$ & 279 & 0.06 \\
\multicolumn{2}{c}{...} & $z \in [0.5,0.6]$ & 396 & 0.08 \\
$z \in [0.4,0.5]$ & 279 & $z \in [0.5,0.6]$ & 396 & 0.23 \\
\tableline
\end{tabular}

\end{table}

\clearpage

\clearpage

\clearpage

\clearpage

\begin{figure}
\epsscale{1.0}
\caption{Observed $(g' - r')$ vs. $(r' - i')$ (ABmag) for the galaxies (red dots) and the templates (black dots). \label{gr_ri_z0p25}}
\end{figure}

\clearpage

\begin{figure}
\epsscale{1.0}
\caption{Observed $(g' - r')$ vs. $(r' - f_{3.6})$ (ABmag) for the galaxies (red dots) and the templates (black dots).\label{gr_r3p6_z0p25}}
\end{figure}

\clearpage

\begin{figure}
\epsscale{1.0}
\caption{Observed $(NUV - g')$ vs. $(g' - r')$ (ABmag) for the galaxies (red dots) and the templates (black dots).\label{nuvg_gr_z0p25}}
\end{figure}

\clearpage

\clearpage

\begin{figure}
\epsscale{0.75}
\plotone{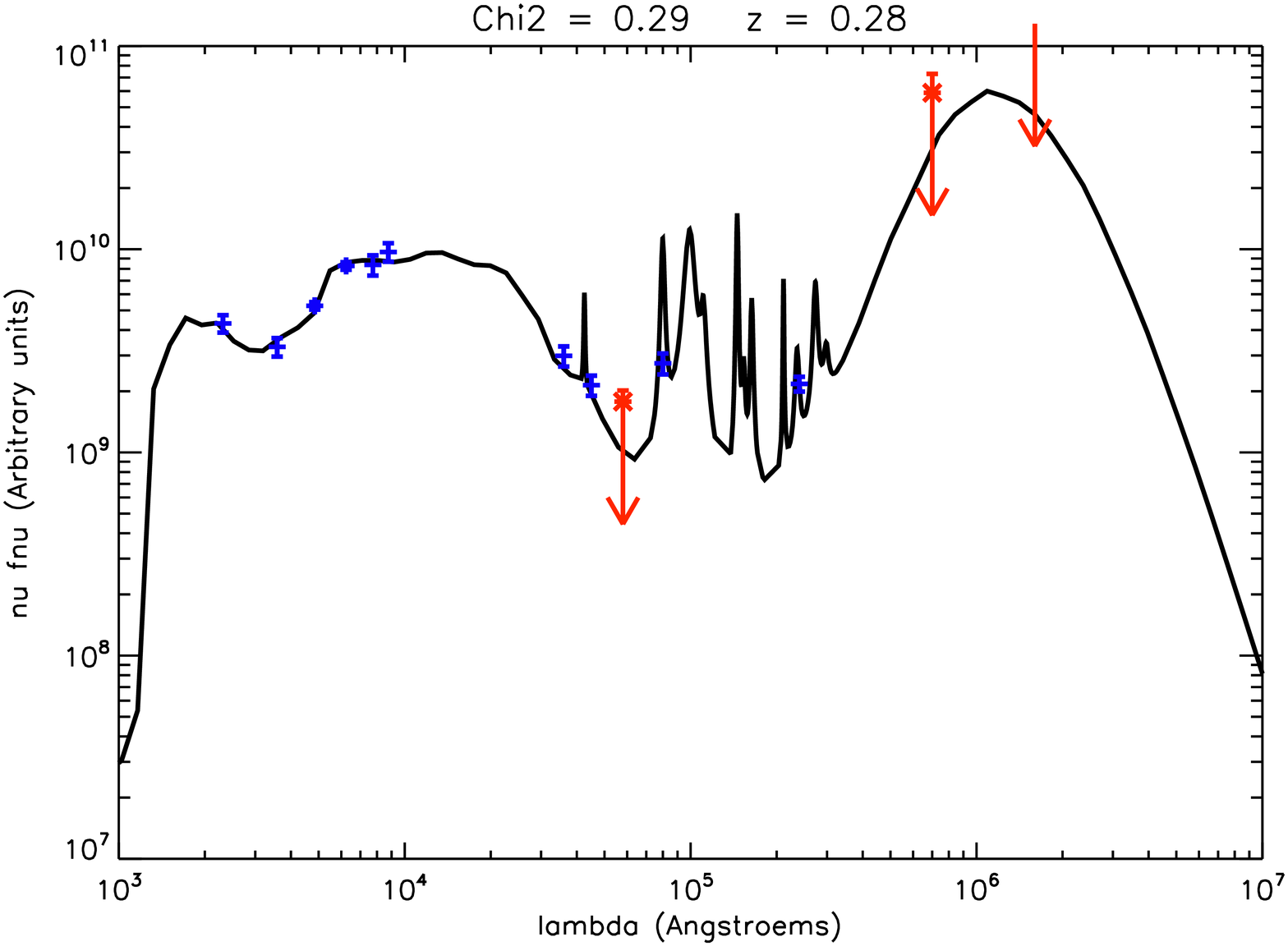}
\plotone{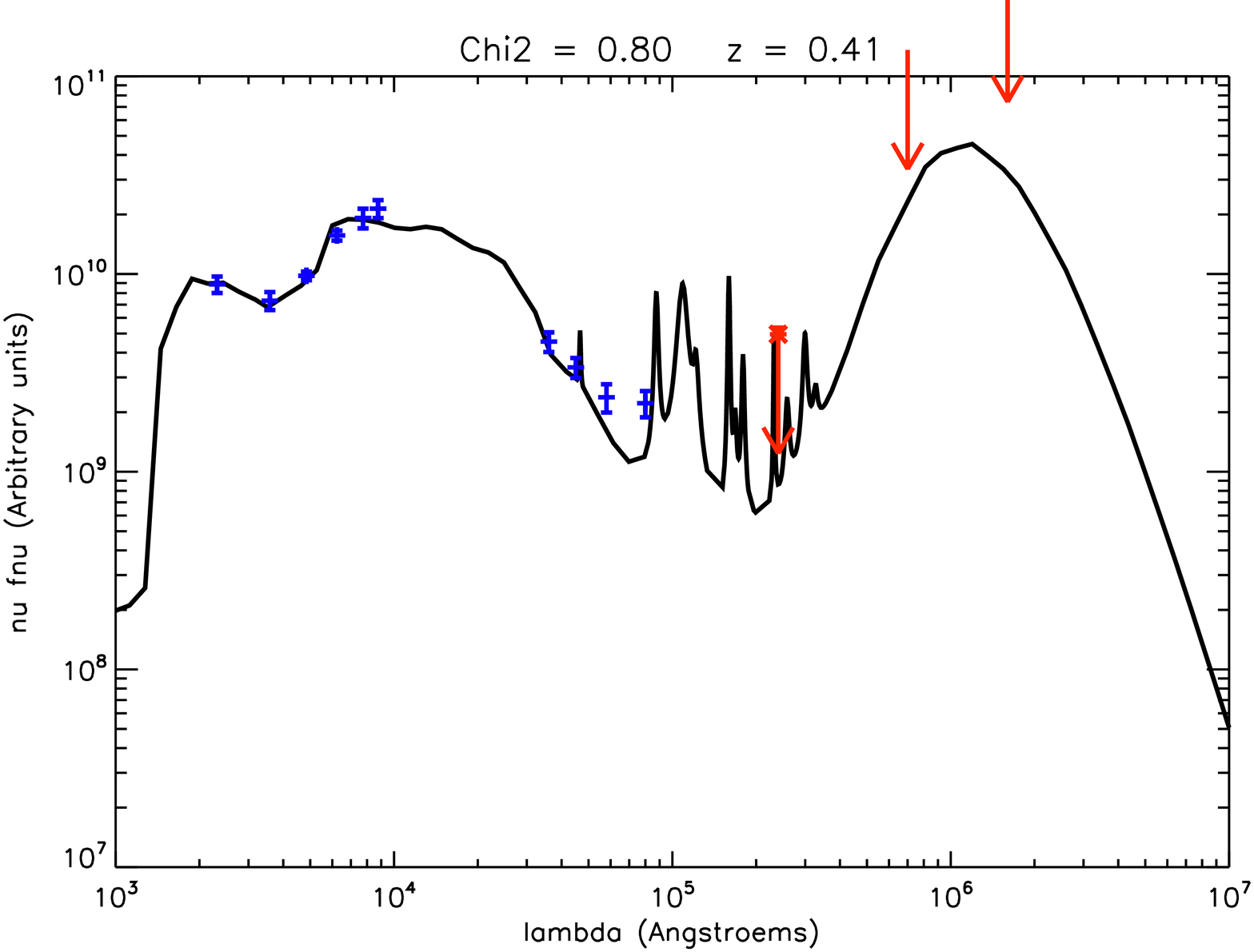}
\caption{Best fits for several galaxies of the sample. Blue symbols correspond to detections. Red arrows correspond to upper limits. Error bars are indicated.\label{plot}}
\end{figure}

\clearpage

\addtocounter{figure}{-1}

\begin{figure}
\epsscale{0.75}
\plotone{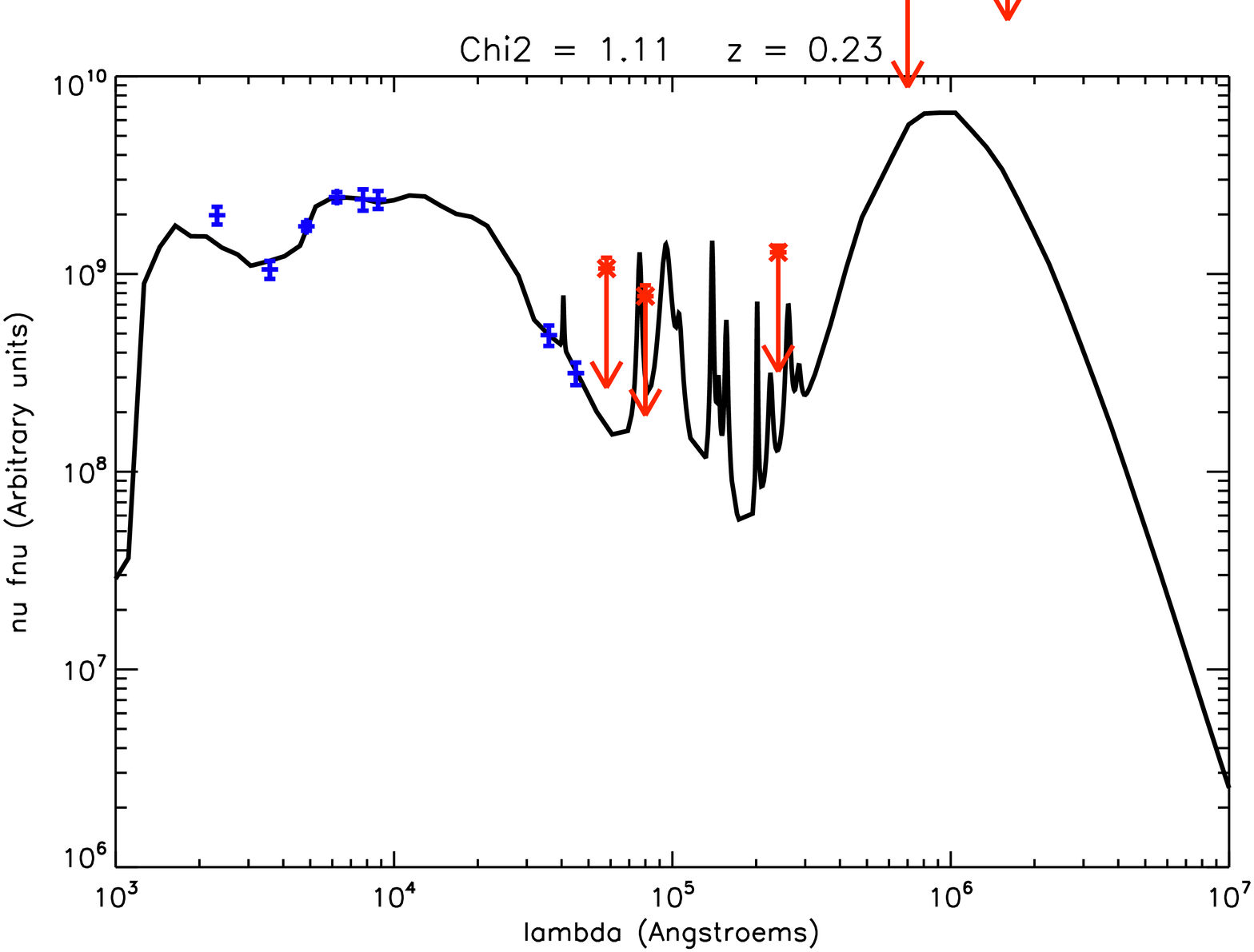}
\plotone{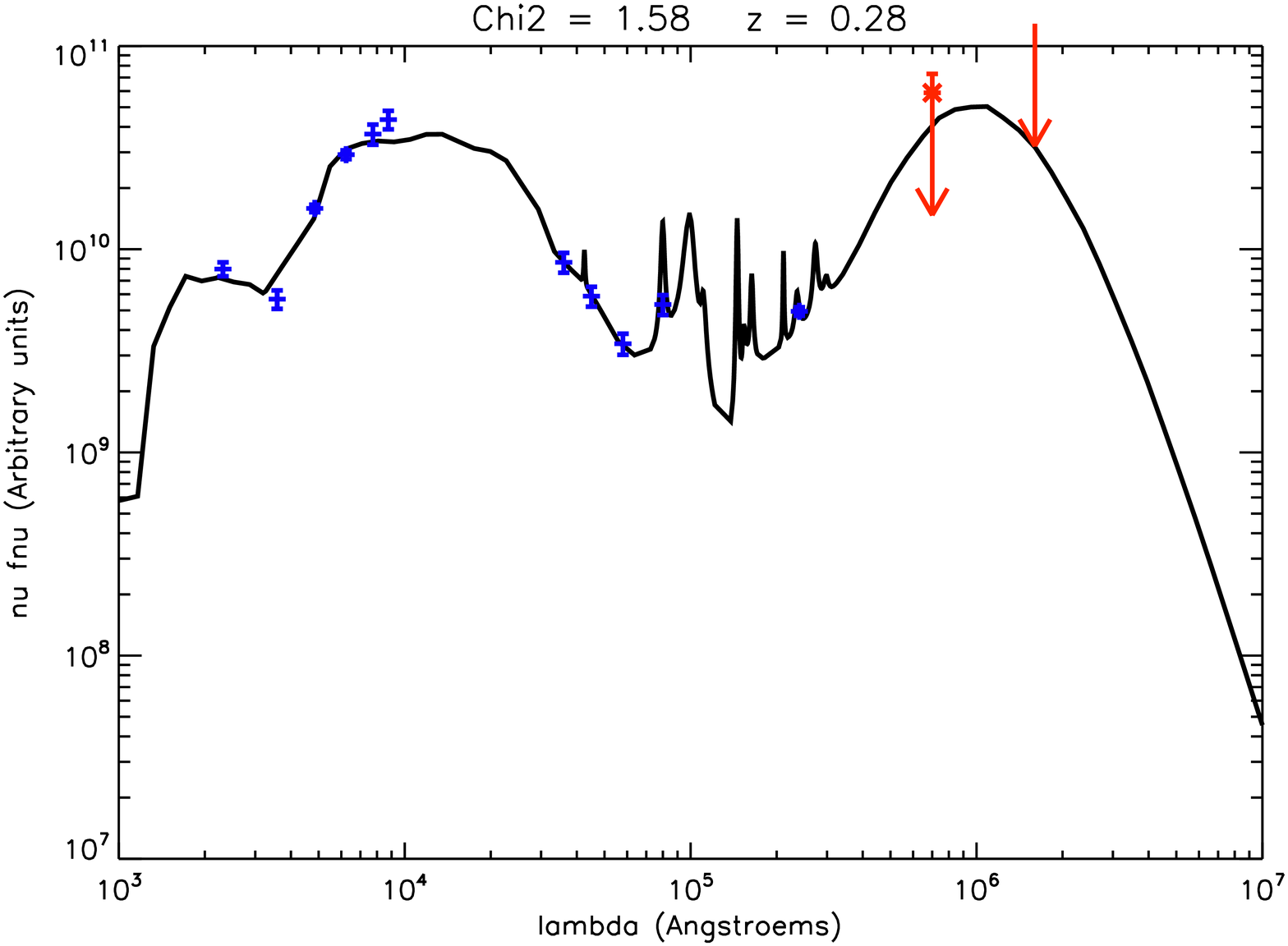}
\caption{Continued.}
\end{figure}

\clearpage

\addtocounter{figure}{-1}

\begin{figure}
\epsscale{0.75}
\plotone{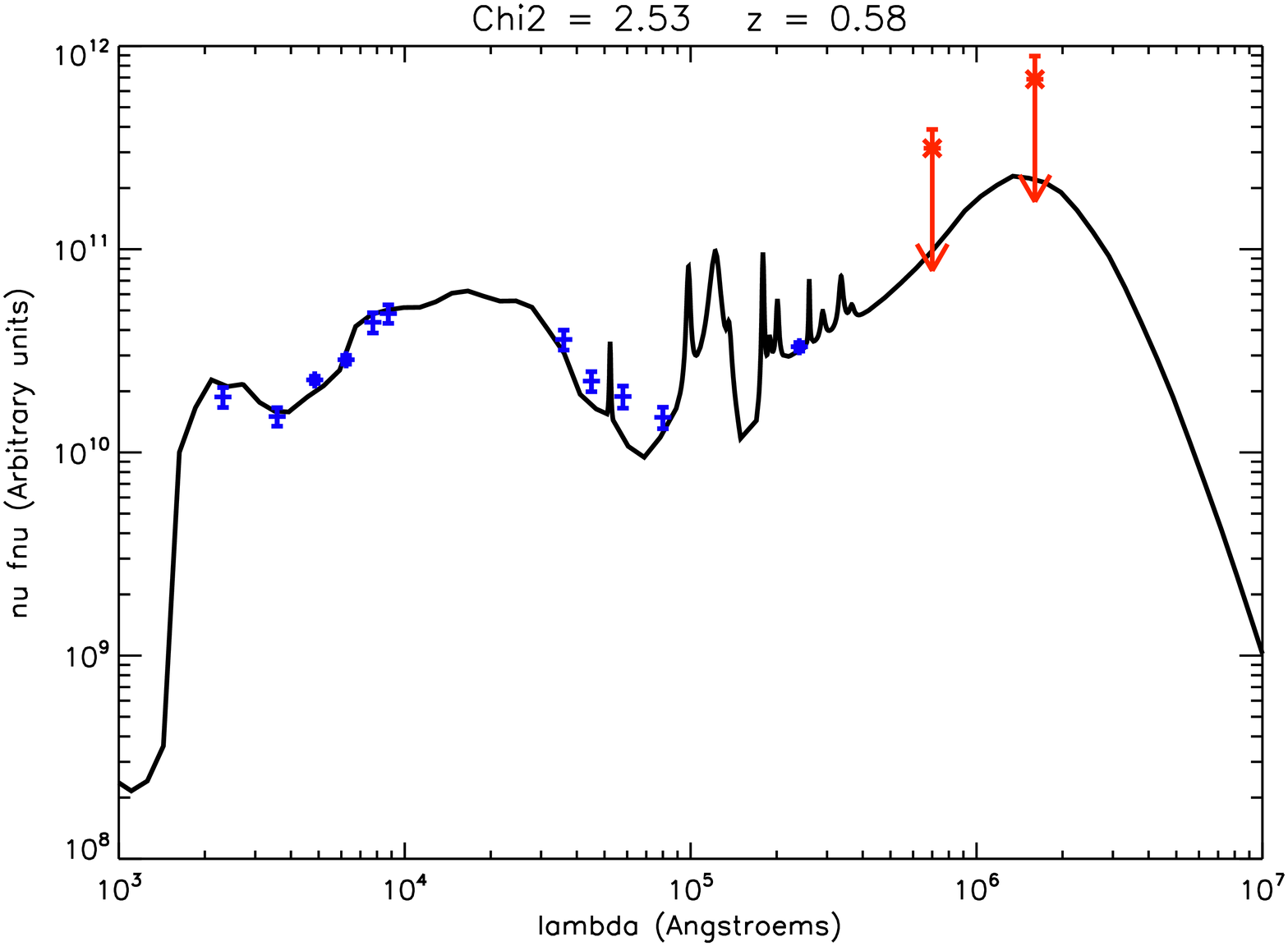}
\plotone{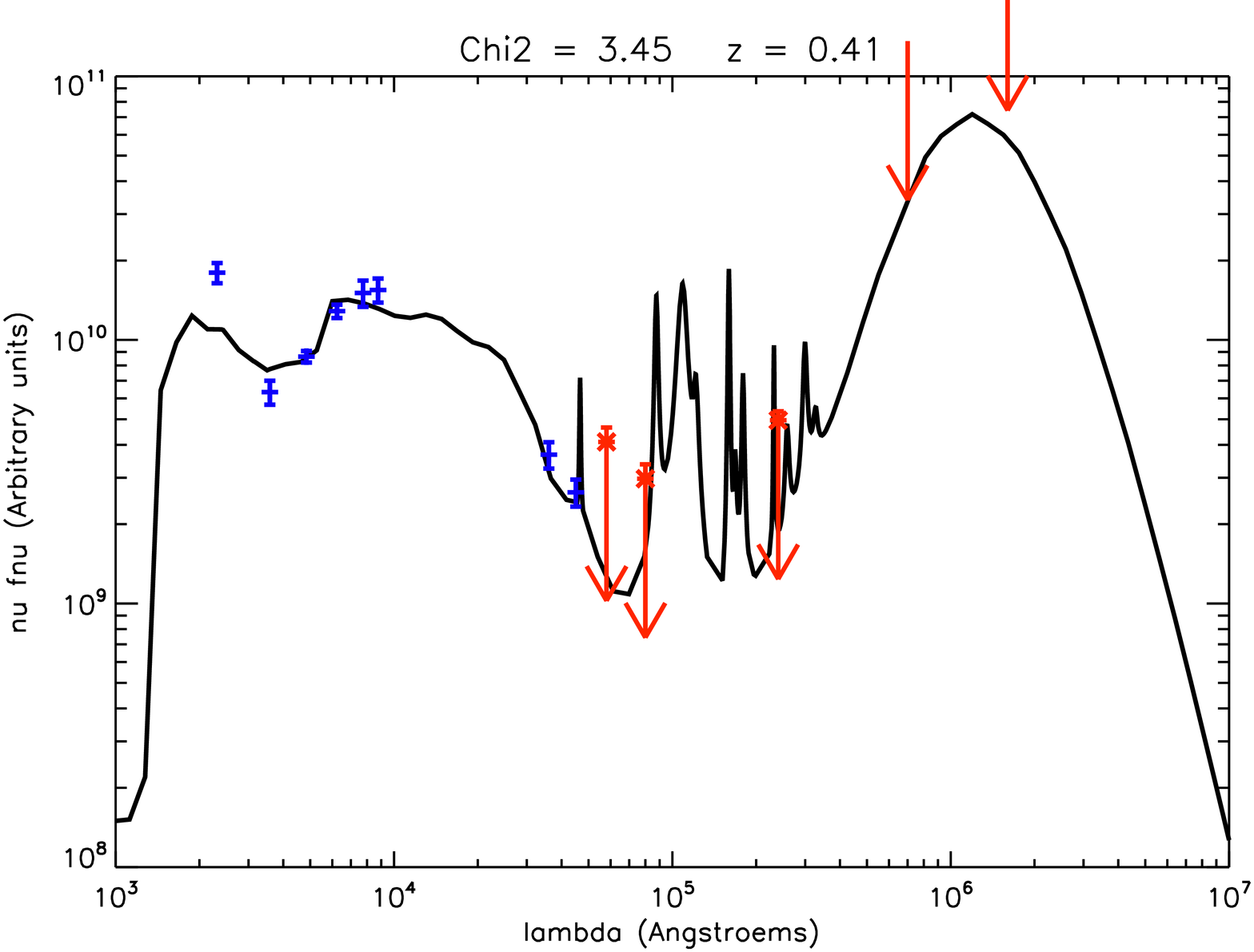}
\caption{Continued.}
\end{figure}

\clearpage

\begin{figure}
\epsscale{1.00}
\plotone{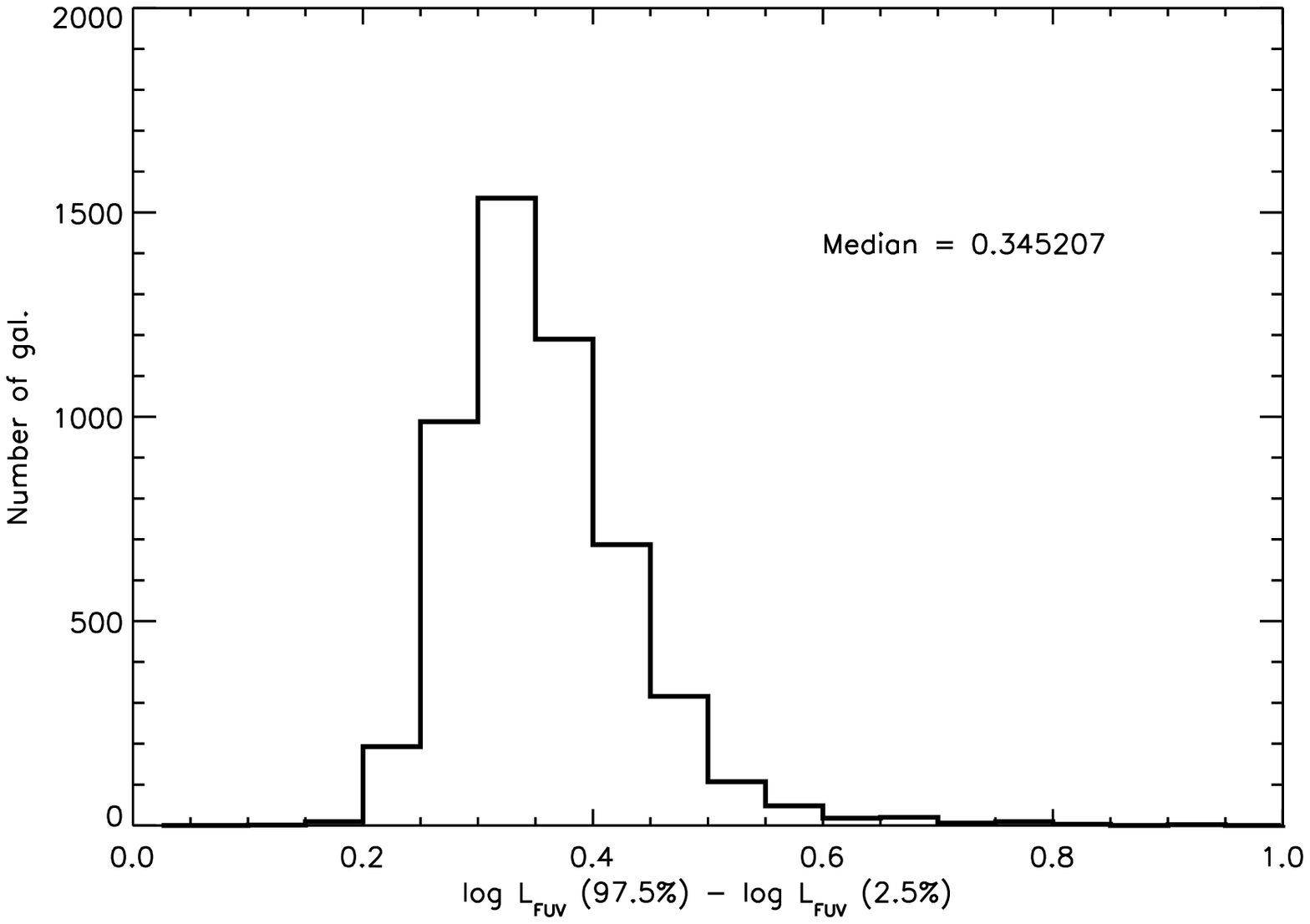}
\caption{Histogram of the width of the 95\% confidence interval of $L_{FUV}$ for our sample galaxies.\label{confi_lfuv}}
\end{figure}

\clearpage

\begin{figure}
\epsscale{1.00}
\plotone{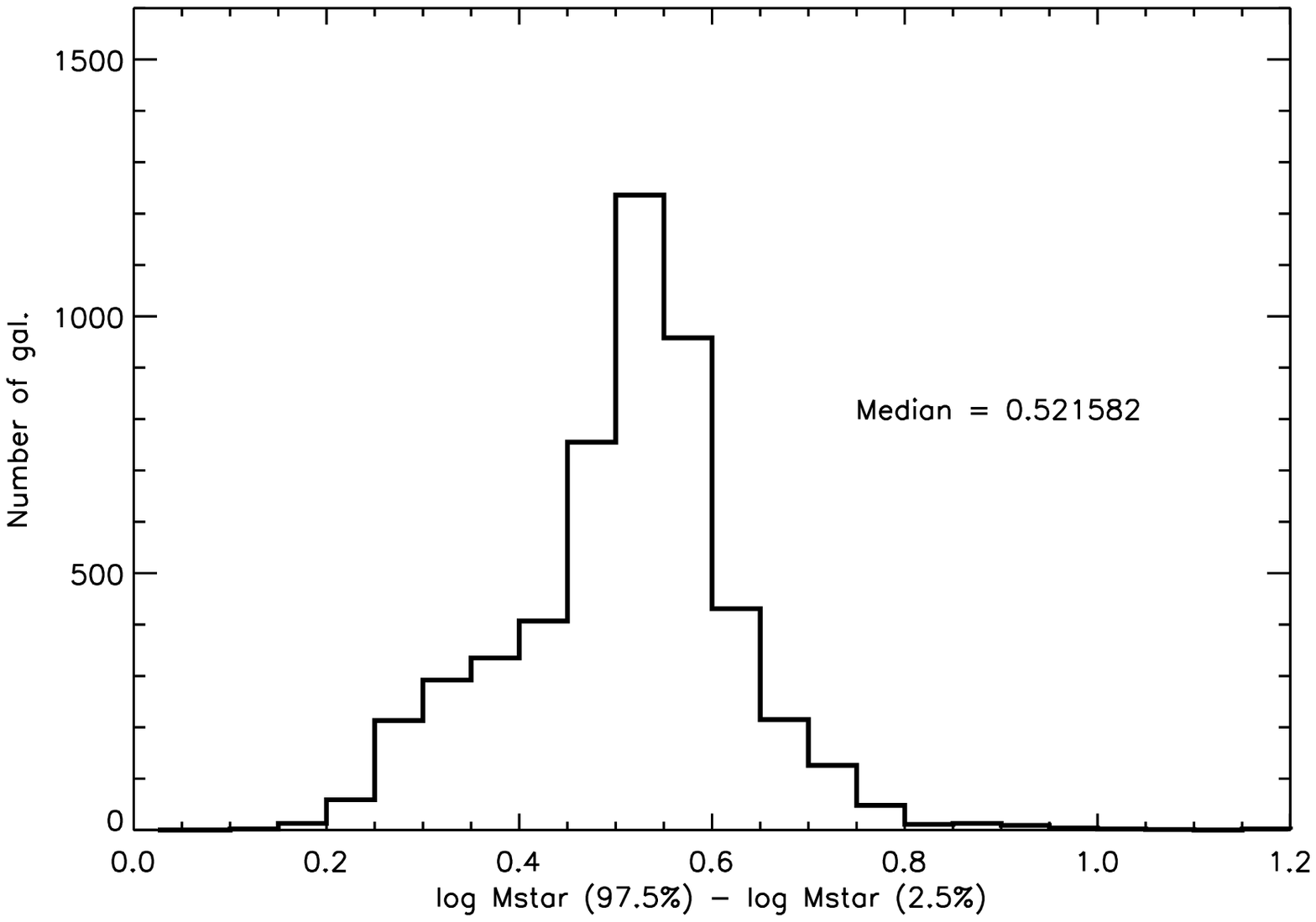}
\caption{Histogram of the width of the 95\% confidence interval of $M*$ for our sample galaxies.\label{confi_mstar}}
\end{figure}

\clearpage

\begin{figure}
\epsscale{1.00}
\plotone{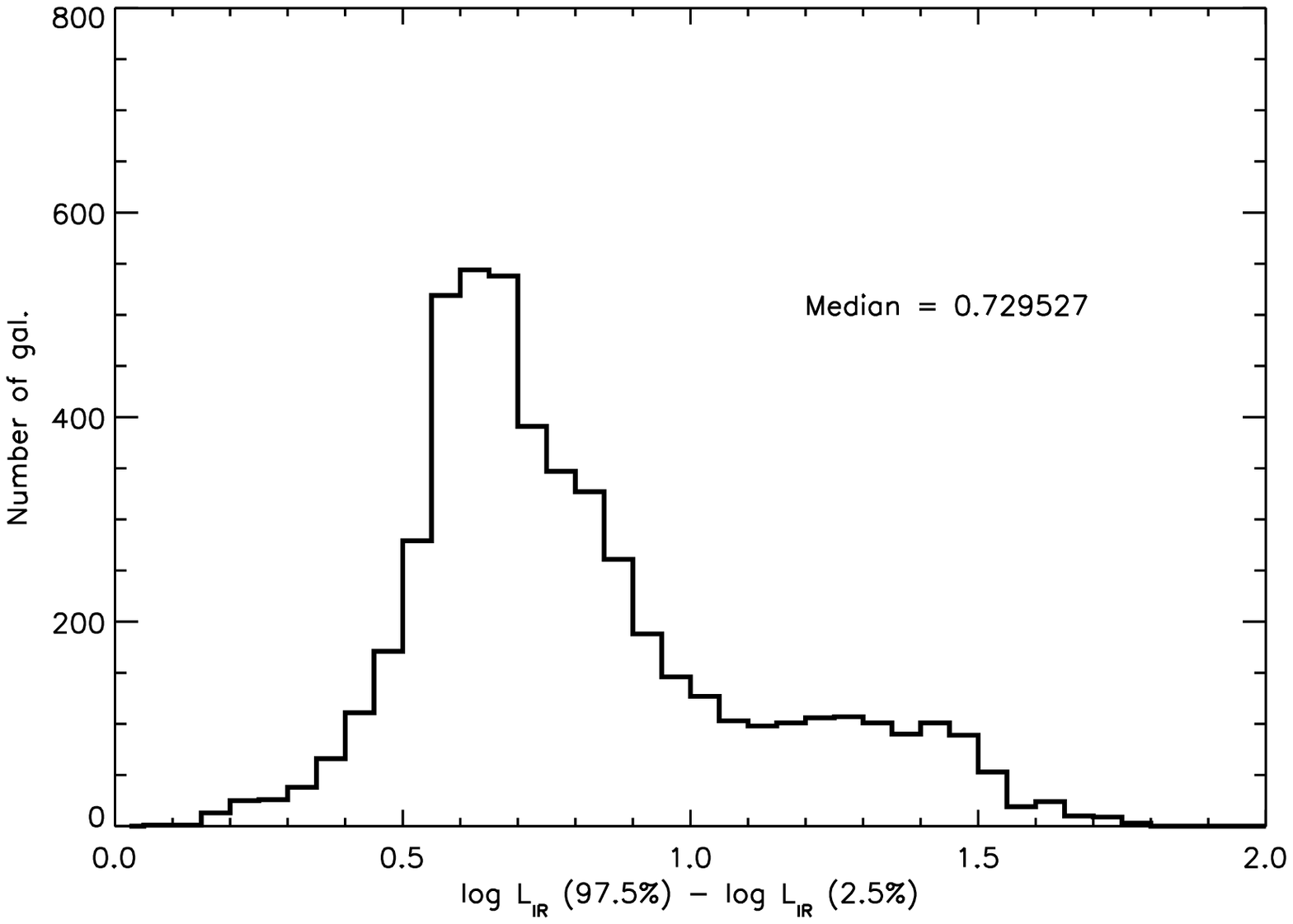}
\caption{Histogram of the width of the 95\% confidence interval of $L_{IR}$ for our sample galaxies.\label{confi_lir}}
\end{figure}

\clearpage

\begin{figure}
\epsscale{1.00}
\caption{$L_{IR}$ from Chary \& Elbaz (2001) models compared to $L_{IR}$ derived from the GRASIL models. Blue (Red) points correspond to galaxies detected at 24$\mu$m (70$\mu$m).
\label{lir_lirelbaz}}
\end{figure}

\clearpage

\begin{figure}
\epsscale{1.00}
\caption{$L_{bol}$ vs. $L_{IR}/L_{FUV}$ for the sample galaxies. 
Grey lines indicate the observational limits above which no galaxies of a given redshift bin are allowed. 
For each panel the corresponding observational limit is shown in bold type. The limiting FUV luminosities were estimated assuming that the limiting value of the observed $m_{NUV} = 22.5~ABmag$ applies to the rest frame FUV luminosity at all redshifts. These limiting values are $L_{FUV} = 10^{9}, 2.52 \times 10^{9}, 4.95 \times 10^{9}, 8.46 \times 10^{9}$ and $1.32 \times 10^{10}~L_{\odot}$ for $z = 0.2, 0.3, 0.4, 0.5$ and 0.6 respectively.
The relation found for UV selected local galaxies by Buat et al. (2006) is shown in red in the top-left panel.
The filled grey regions are the areas (free of observational biases) over which we have estimated the average $L_{IR}/L_{FUV}$ for the different redshift bins. \label{lbol_lirlfuv}}
\end{figure}

\clearpage

\begin{figure}
\epsscale{0.80}
\plotone{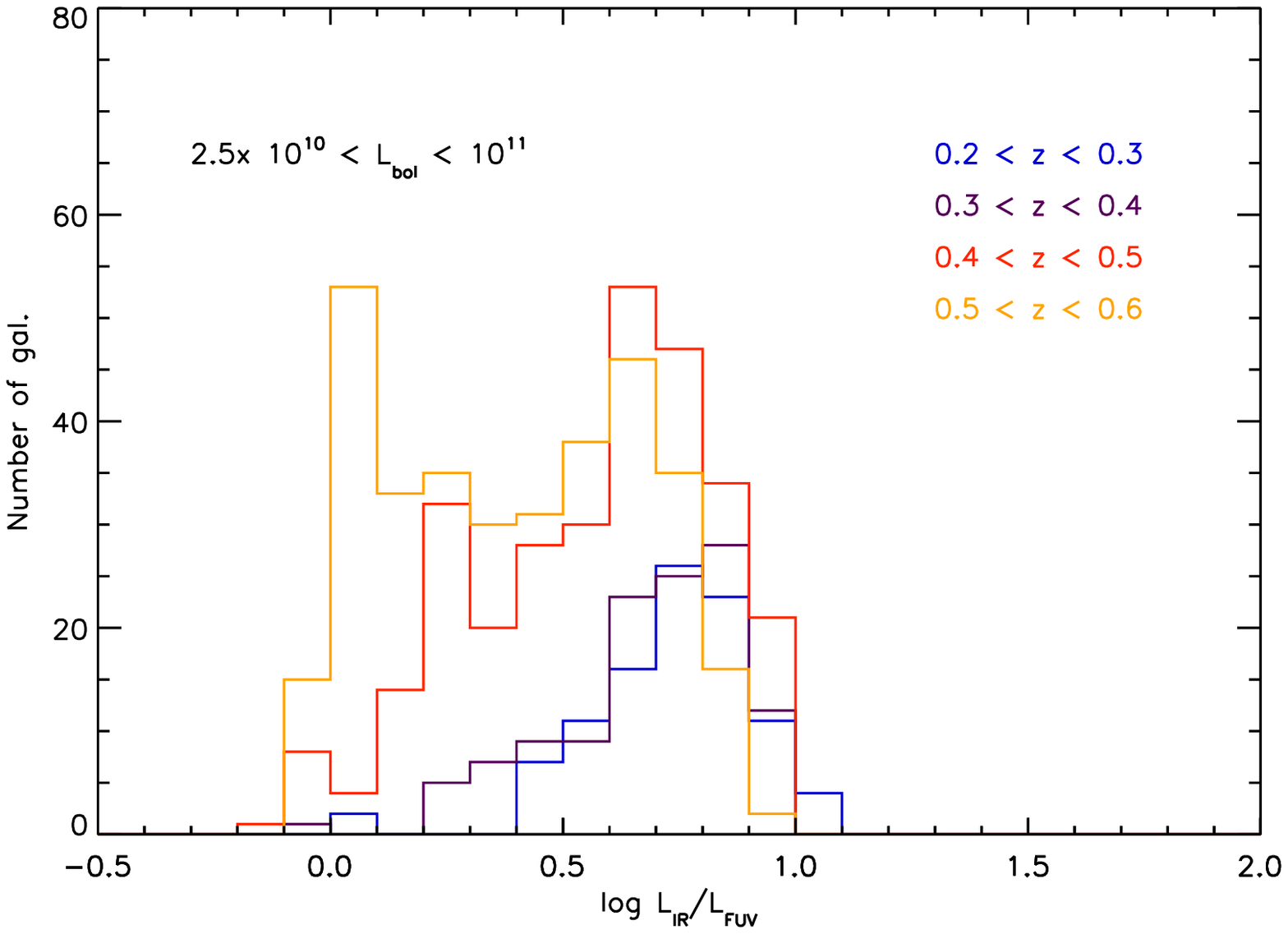}\plotone{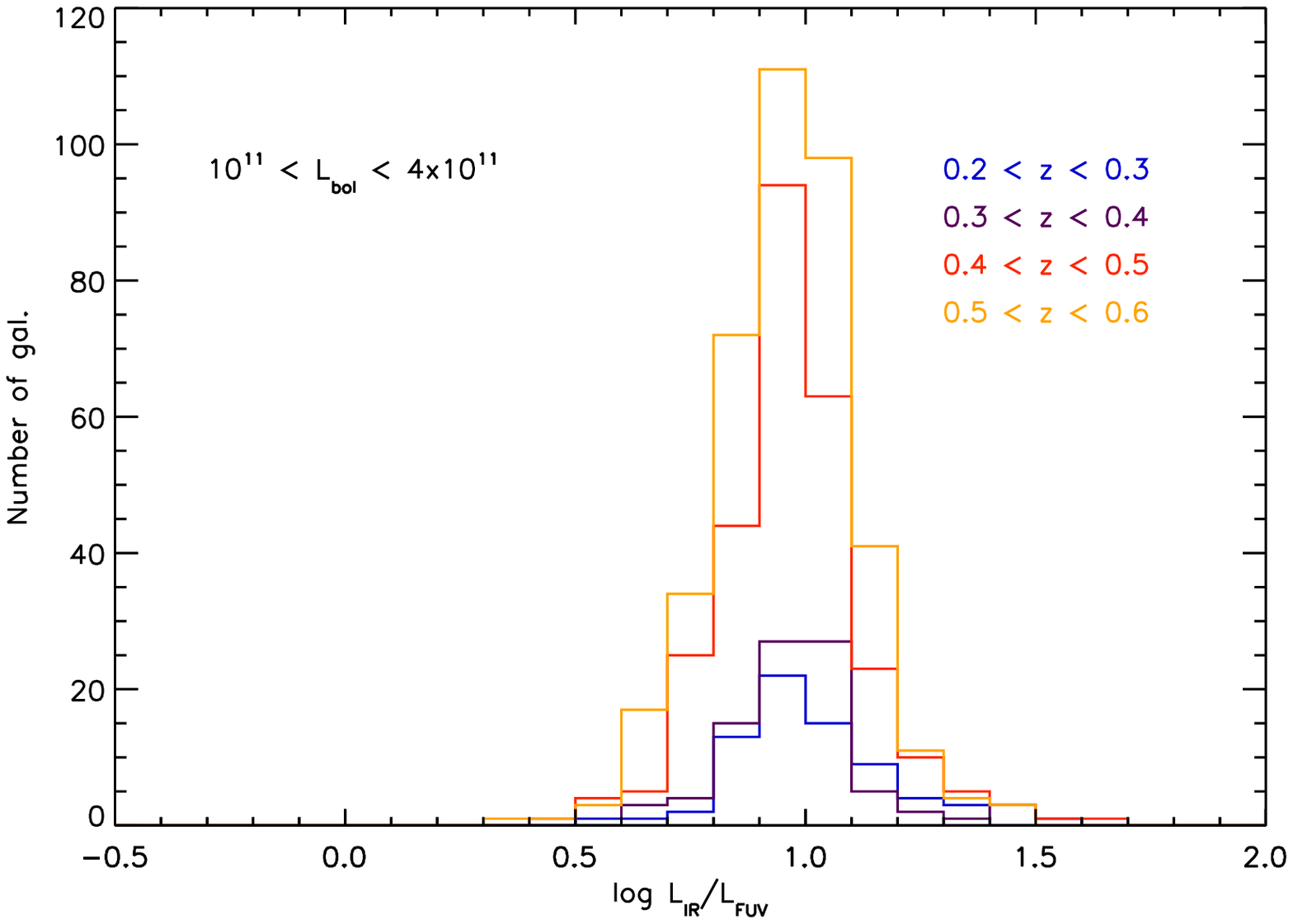}
\caption{Distribution of $L_{IR}/L_{FUV}$ for regions (a) (upper plot) and (b) (lower plot) of Figure~\ref{lbol_lirlfuv}. \label{histo_lbol}}
\end{figure}

\clearpage

\begin{figure}
\epsscale{1.00}
\caption{$M*$ vs. $L_{IR}/L_{FUV}$ for the sample galaxies. 
\label{mstar_lirlfuv}}
\end{figure}

\clearpage

\begin{figure}
\epsscale{1.00}
\plotone{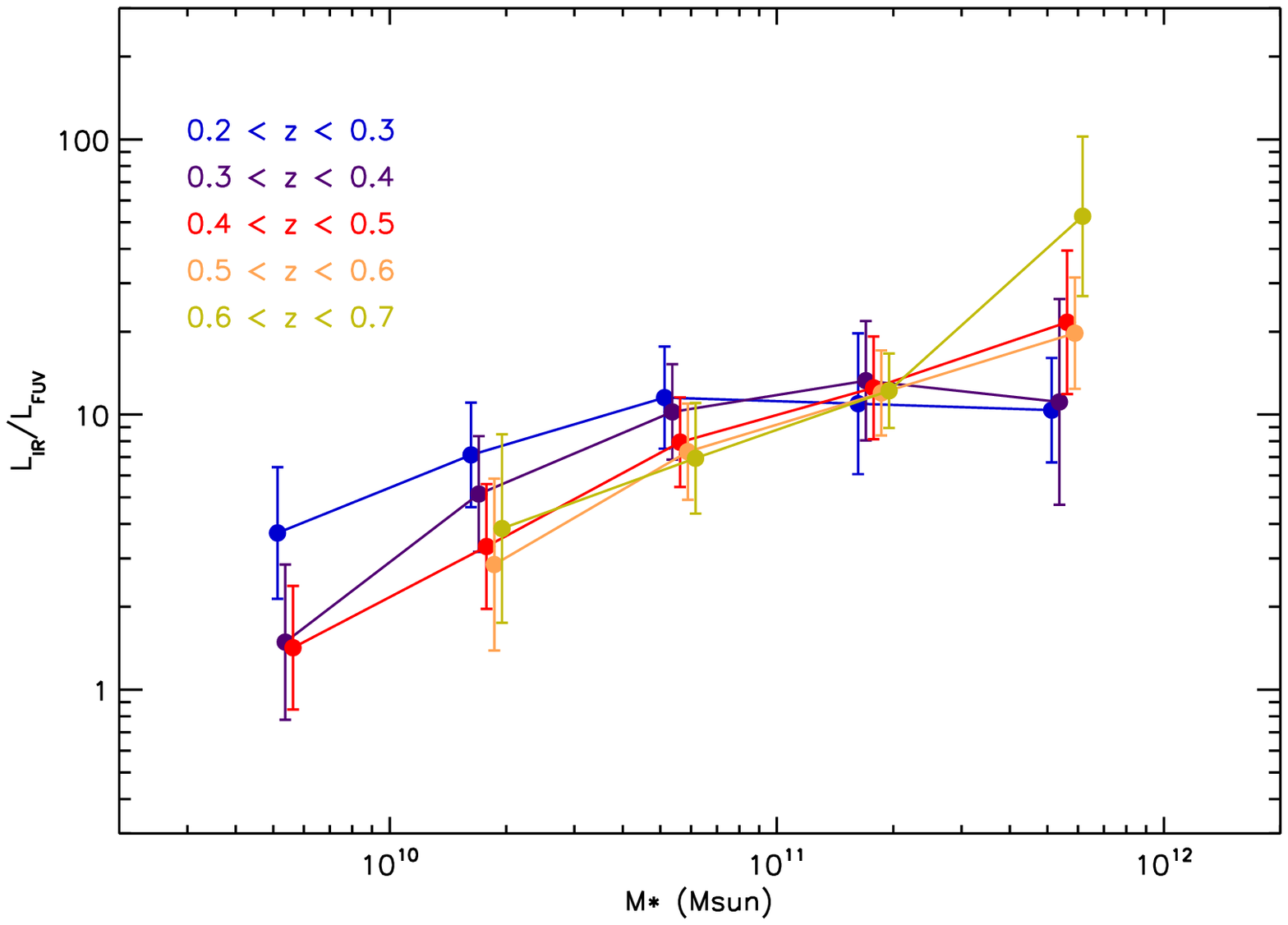}
\caption{Volume weighted average $L_{IR}/L_{FUV}$ vs. M* for the different redshift intervals. Error bars correspond to (volume weighted) standard deviations. The mass intervals correspond to: $M* \in [10^{9.5},10^{10}], [10^{10},10^{10.5}], [10^{10.5},10^{11}], [10^{11},10^{11.5}], [10^{11.5},10^{12}]$. The points have been slightly shifted horizontally in order to avoid confusion. 
\label{mstar_lirlfuv_mean}}
\end{figure}

\clearpage

\begin{figure}
\epsscale{1.00}
\caption{M* vs. $\nu$ for the sample galaxies. For each galaxy, the value of $\nu$ is the average over all the models weighted by $\exp(- \chi^{2}/2)$.
\label{mstar_nu}}
\end{figure}

\clearpage

\begin{figure}
\epsscale{1.00}
\plotone{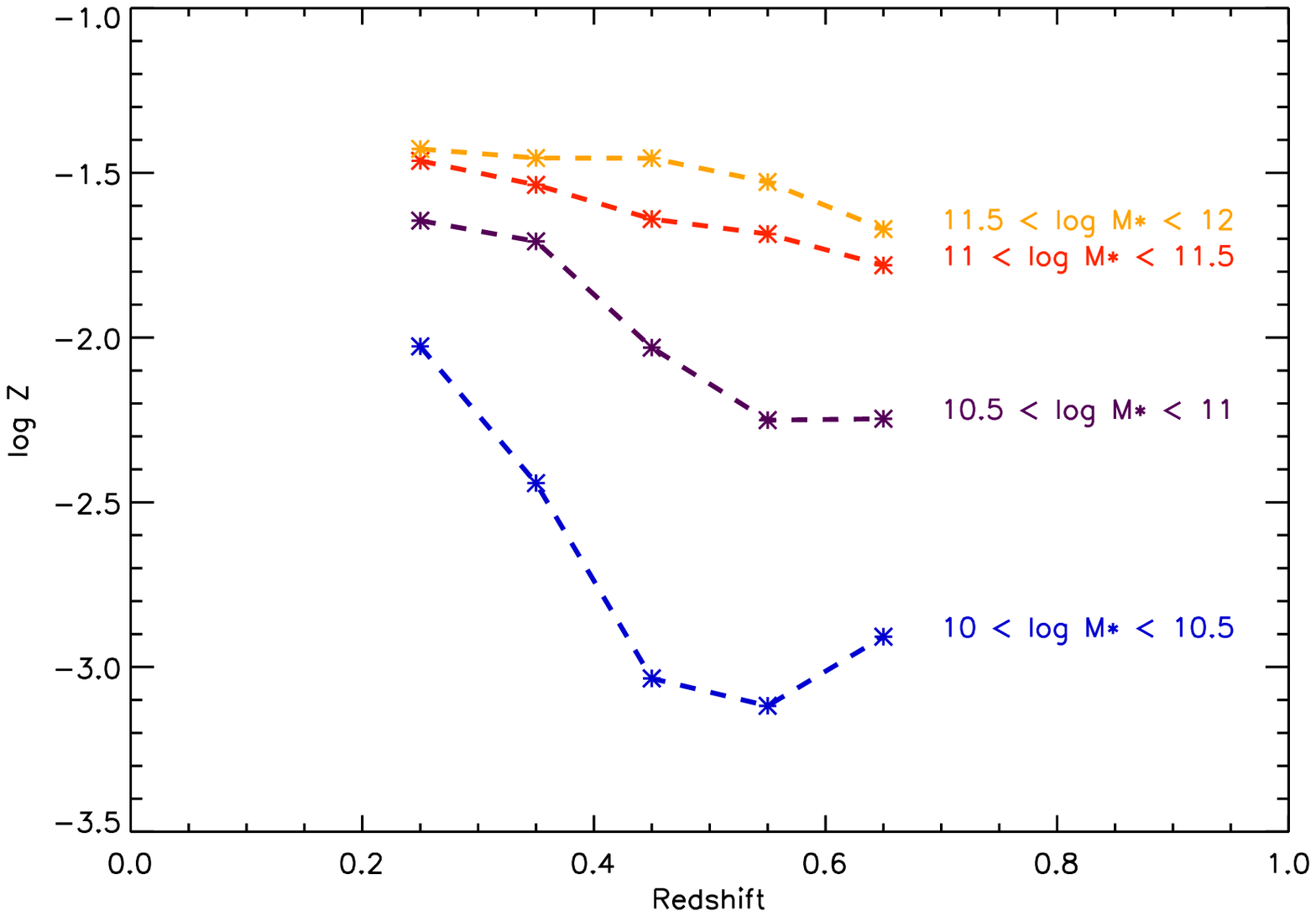}
\caption{Redshift vs. weighted averaged metallicity (in solar units) for different stellar mass.
\label{zeta_mass}}
\end{figure}

\clearpage

\begin{figure}
\epsscale{1.00}
\caption{$M*$ vs. $SFR_{8}/M*$ for the sample galaxies. 
The grey lines correspond to the (approximate) observational limits for each redshift bin. 
In each panel the observational limit corresponding to the indicated redshift bin is shown in bold type.
The average relation found for UV selected local galaxies by Buat et al. (2006) is represented by a dashed cyan line. 
The horizontal lines represent the required SSFR to double a galaxy's mass between each redshift epoch and the present.\label{mstar_ssfr}}
\end{figure}

\clearpage

\begin{figure}
\epsscale{1.00}
\plotone{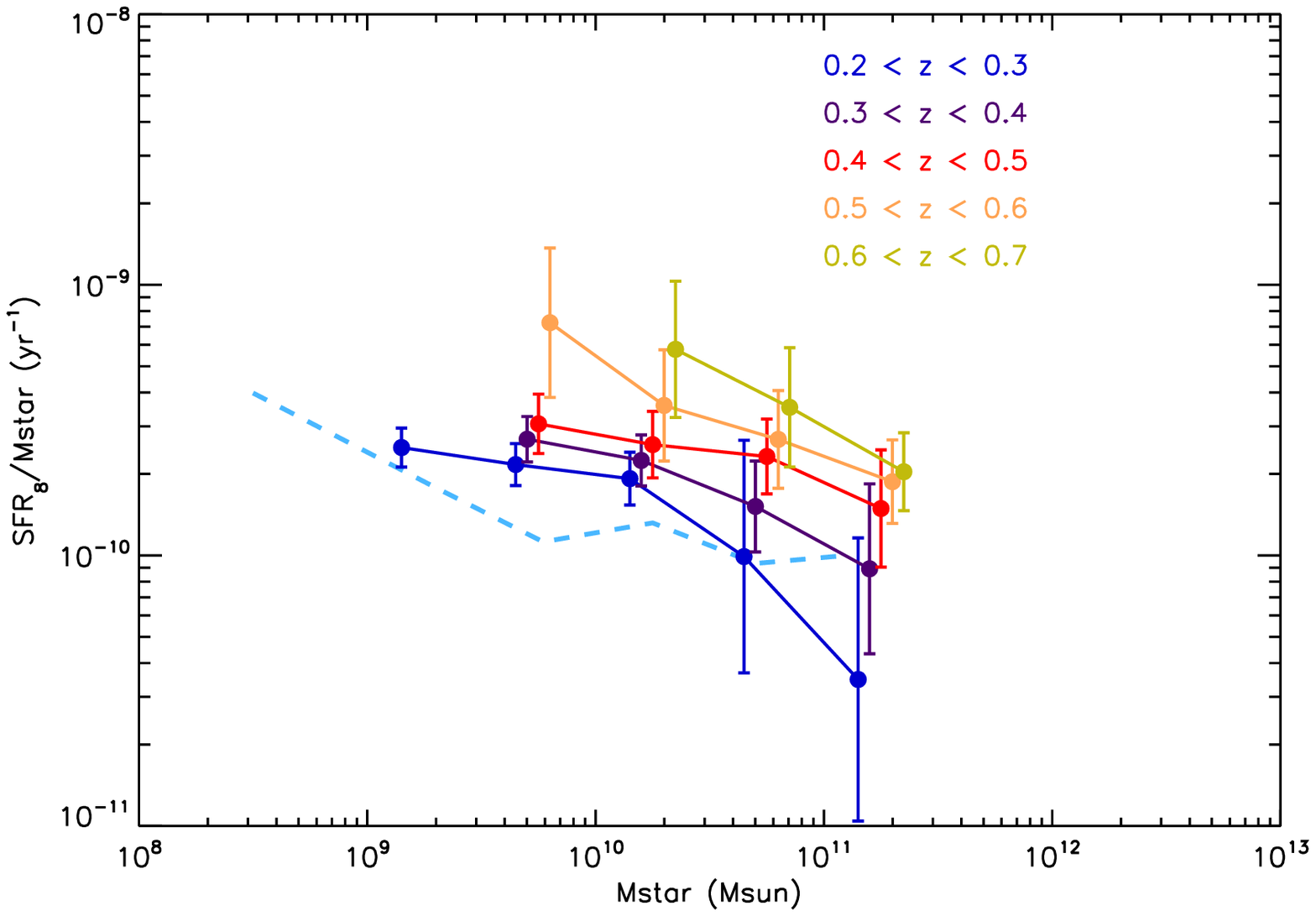}
\caption{Volume weighted average values of SSFR as a function of $M*$.
The mass intervals are: $\log M* \in [9,9.5], [9.5,10], [10,10.5], [10.5,11], [11,11.5]$. 
Points were slightly shifted horizontally by different quantities in order to avoid overlapping. \label{mstar_ssfr_volumen}}
\end{figure}

\clearpage

\begin{figure}
\epsscale{1.00}
\caption{$M*$ vs . $(U - V)$ rest frame for our sample galaxies. Solid lines separate the red and blue sequences at each redshift (from Bell et al. 2007). \label{secuencia}}
\end{figure}

\end{document}